\documentclass{LMCS}

\input{macros.sty}
\usepackage{enumerate,hyperref}

\def\doi{6 (4:7) 2010}
\lmcsheading%
{\doi}
{1--31}
{}
{}
{Nov.~16, 2009}
{Dec.~18, 2010}
{}

\begin{document}

\title[Bounded Linear Logic, Revisited]{Bounded Linear Logic, Revisited\rsuper*}
\titlecomment{{\lsuper*}This is a revised and extended version of a
  paper with the same title appeared in the Proceedings of the Ninth
  International Conference on Typed Lambda Calculi and Applications}

\author[U.~Dal Lago]{Ugo Dal Lago\rsuper a}
\address{{\lsuper a}Dipartimento di Scienze dell'Informazione, Universit\`a di Bologna} 
\email{dallago@cs.unibo.it}
\author[M.~Hofmann]{Martin Hofmann\rsuper b}
\address{{\lsuper b}Institut f\"u{}r Informatik, LMU M\"u{}nchen}
\email{mhofmann@informatik.uni-muenchen.de}


\begin{abstract}
  We present \QBAL, an extension of Girard, Scedrov and Scott's bounded
  linear logic. The main novelty of the system is
  the possibility of quantifying over resource variables. This
  generalization makes bounded linear logic considerably more flexible, while
  preserving soundness and completeness for polynomial time. In particular, we provide
  compositional embeddings of Leivant's \RRW\ and Hofmann's \LFPL\
  into \QBAL.
\end{abstract}
\maketitle

\section{Introduction}
\noindent After two decades from the pioneering works that started it~\cite{Bellantoni92,Leivant91,Leivant93}, 
implicit computational complexity is now an active research area at the intersection of
mathematical logic and computer science. Its aim is the study of machine-free characterizations
of complexity classes. The correspondence between an ICC system and a complexity class
holds only \emph{extensionally}, i.e., the class of functions (or problems) which are
representable in the system equals the complexity class. Usually, the 
system is a fragment or subsystem of a larger programming language
or logical system, the \emph{base system}, in which other functions besides the ones in the complexity class
can be represented. Sometimes, one of the two
inclusions is shown by proving that any program (or proof) can be reduced with a bounded
amount of resources; in this case, we say that the system is intensionally sound. On the other
hand, ICC systems are very far from being \emph{intensionally complete}: there are
many programs (or proofs) in the base system which are not in the ICC system,
even if they can be evaluated
 with the prescribed complexity bounds. Observe that this does not
contradict extensional completeness, since many different programs or proofs compute
the same function.

Of course, a system that captures all and only the programs of the
base system running within a prescribed complexity bound will in all
but trivial cases (e.g., empty base system) fail to be recursively enumerable. 
Thus, in practice, one strives to improve intensional
expressivity by capturing important classes of examples and patterns.

An obstacle towards applying ICC characterizations of complexity classes
to programming language theory is their poor intensional expressive power: most 
ICC systems do not capture natural programs and therefore are not useful in practice.
This problem has been already considered in the literature. Some papers try
to address the poor intensional expressive power of ICC systems by defining new
languages or logics allowing to program in ways which are not allowed
in existing ICC systems. This includes quasi-interpretations~\cite{Marion00}
and \LFPL, by the second author~\cite{Hofmann03}. Other papers analyze the intensional expressive
power of existing systems either by studying necessary conditions on captured programs
or, more frequently, by studying relations between existing
ICC systems. One nice example is Murawski and Ong's paper~\cite{Murawski04}, in which the
authors prove that a subsystem \BCmin\ of Bellantoni and Cook's function algebra \BC~\cite{Bellantoni92} can 
be embedded into light affine logic~\cite{Asperti02} and that the embedding cannot be
extended to the whole \BC.
In this work, we somehow combine the two approaches, by showing that:
\begin{enumerate}[$\bullet$]
\item
  A new logical system, called Quantified Bounded Affine Logic (\QBAL\ for short)
  can be defined as a generalization of
  Girard, Scedrov and Scott's bounded linear logic (\BLL,~\cite{Girard92}),
  itself the first characterization of polynomial time computable functions as a 
  fragment of Girard's linear logic~\cite{Girard87}.
\item
  \QBAL\ is intensionally at least as expressive as two heterogeneous, 
  existing systems, namely Leivant's \RRW~\cite{Leivant93} and \LFPL. 
\end{enumerate}

Bounded linear logic has received relatively little attention in the past~\cite{Hofmann04,Schoepp07}. This
is mainly due to its syntax, which is more involved than the one of other complexity-related
fragments of linear logic appeared more recently~\cite{Girard98,Lafont04,Danos03}. 
In bounded linear logic, polynomials are part of the syntax and, as a
consequence, computation time is controlled \emph{explicitly}; in other words, \BLL\ cannot
be claimed to be a truly implicit characterization of polynomial time. 
Moreover, it seems that \BLL\ is not expressive enough
to embed any existing ICC system corresponding to polynomial time (except
Lafont's \SLL~\cite{Lafont04}, which anyway was conceived as a very
small fragment of \BLL). 

\QBAL\ is obtained by endowing \BLL\ with bounded quantification on resource 
variables. In other words, formulas of \QBAL\ includes the ones of \BLL, 
plus formulas like $\exists
x:\{x\leq y^2\}.A$ or $\forall x,y:\{x\leq z,y\leq z^3\}.B$. Rules governing
bounded quantification can be easily added to \BLL, preserving its
good properties: \QBAL\ is still a characterization of the polytime functions
in an extensional sense. Bounded quantification on resource variables, on
the other hand, has tremendous consequences from an intensional point of view: 
both \RRW\ and \LFPL\ can be compositionally embedded into \QBAL. This means,
arguably, that programs in either \RRW\ or \LFPL\ can
be rewritten in \QBAL\ without major changes, i.e., by mimicking their 
syntactic structure. Similar results are unlikely to hold for \BLL, as
argued in sections~\ref{sect:emlfpl} and~\ref{sect:emrrw} below.

Logical systems like \QBAL\ or \BLL\ cannot be considered as practical
programming languages, although proofs can be interpreted as programs in
the sense of Curry and Howard: the syntax is too complicated and the potential
programmer would have to provide quantitative information in the form of
polynomials while writing programs. On the other hand, considering \BLL\ or
\QBAL\ as type systems for the (linear) lambda calculus is interesting, although
type inference would be undecidable in general. In this paper, we advocate
the usefulness of \QBAL\ as an intermediate language in which to prove soundness
results about other ICC systems.

For all these reasons, \QBAL\ is \emph{not} just another system capturing polynomial
time computable functions.

The rest of this paper is organized as follows:
\begin{enumerate}[$\bullet$]
\item
  In Section \ref{sect:syntax} the syntax of \QBAL\ is introduced, and some
  interesting properties of the system are proved, together with some
  examples on how to write programs as proofs of \QBAL.
\item
  Section \ref{sect:sts} introduces a set-theoretic semantics
  for \QBAL, which is exploited in Section \ref{sect:qpt} to
  show polytime soundness of the system.
\item
  Section \ref{sect:oce} contains some informal discussion about
  compositional embeddings.
\item
  Section \ref{sect:emlfpl} and \ref{sect:emrrw} present compositional
  embeddings of \LFPL\ and \RRW\ into \QBAL, respectively.
\end{enumerate}
\section{Syntax}\label{sect:syntax}
\noindent In this section, we present the syntax of \QBAL, together with some of its main properties.
In the following, we adhere to the notation adopted in the
relevant literature on \BLL~\cite{Girard92,Hofmann04}.
\subsection{Resource Polynomials and Constraints}
Polynomials appears explicitly in the formulas of \QBAL, exactly as in \BLL. A 
specific notation for polynomials was introduced in \cite{Girard92}, and will be
adopted here. In the following, $\N$ is the set of natural numbers. Sometimes,
we use the vector notation $\overline{x}$, which stands for the sequence $x_1,\ldots,x_n$,
where $n$ is assumed to be known from the context.
\begin{defi}\hfill
\begin{enumerate}[$\bullet$]
  \item
    Given a set $X$ of \emph{resource variables}, a 
    \emph{resource monomial over $X$} is any finite product of 
    binomial coefficients
    $$
    \prod_{i=1}^m\binom{x_{i}}{n_{i}}
    $$
    where the resource variables $x_1,\ldots,x_m$ are pairwise distinct
    and $n_1,\ldots,n_m$ are natural numbers.
  \item
    A \emph{resource polynomial over $X$} is any finite sum of
    monomials over $X$. $\FV{p}$ denotes the set of resource variables in a resource polynomial $p$.
\end{enumerate}
\end{defi}

\noindent Resource polynomials are notations for polynomials with rational coefficients. However, by construction,
every resource polynomial maps natural numbers to natural numbers. An example
of a resource polynomial on $\{x\}$ (actually a monomial) is
$$
\binom{x}{2}=\frac{x\cdot(x-1)}{2!}=\frac{1}{2}\cdot x\cdot (x-1).
$$
Resource polynomials satisfy some nice closure properties:
\begin{lem}
All constant functions and the identity are resource polynomials. Moreover,
resource polynomials are closed under binary sums, binary products,
composition, and bounded sums.
\end{lem}
\proof
Every constant function $n$ is simply the resource polynomial
$$
\underbrace{1+\ldots+1}_{\mbox{$n$ times}}=\underbrace{\binom{x}{0}+\ldots+\binom{x}{0}}_{\mbox{$n$ times}}.
$$
The identity is the resource polynomial 
$$
\binom{x}{1}.
$$
Closure under binary sums is trivial. To prove closure under products,
it suffices to show that the product of two monomials (on the
same variable) is a resource polynomial, but this boils down to show that the product
of two binomial coefficients $\binom{x}{n}$ and $\binom{x}{m}$
can be expressed itself as a resource polynomial. Actually, if $m\leq n$, then
$$
\binom{x}{m}\cdot\binom{x}{n}=\sum_{k=0}^m \binom{m+n-k}{k,m-k,n-k}\cdot\binom{x}{m+n-k}
$$
where
$$
\binom{m+n-k}{k,m-k,n-k}
$$
is a multinomial coefficient. The bounded sum
$$
\sum_{x<y}p
$$
where $p$ is a resource polynomial not mentioning the variable $y$ (but possibly mentioning $x$)
can be formed by observing that
$$
\sum_{k=0}^n\binom{j}{k}=\binom{n+1}{k+1}.
$$
Closure by composition can be proved similarly.\qed

As a consequence, every polynomial with natural number coefficients is a resource
polynomial. The main reason why resource polynomials were originally chosen as a notation for
polynomials in \BLL~\cite{Girard92} was closure under bounded sums, a property which is not true in more
traditional notation schemes. We follow the original paper here.

Already in \BLL, an order relation on resource polynomials is an essential ingredient
in defining the syntax of formulas and proof. In \QBAL, the notion is even more important: two
polynomials can be compared unconditionally or with an implicit assumption in the form of a
set of constraints.
\begin{defi}[Constraints]\hfill
\begin{enumerate}[$\bullet$]
\item
  A \emph{constraint} is an inequality in the form $p\leq q$, where $p$ and $q$ are
  resource polynomials. A constraint $p\leq q$ \emph{holds} (and we write $\models p\leq q$)
  if it is true in the standard model. The expression $p<q$ stands for the constraint $p+1\leq q$.
  Variables in $p$ appear negatively, while those in $q$ appear positively in every constraint
  $p\leq q$.
\item
  A \emph{constraint set} is a finite set of constraints.
  Constraint sets are denoted with letters like $\consetone$, $\consettwo$ or
  $\consetthree$. A constraint $p\leq q$ is a consequence of a constraint
  set $\consetone$ (and we write $\consetone\models p\leq q$) 
  if $p\leq q$ is logical consequence of $\consetone$. For every constraint sets
  $\consetone$ and $\consettwo$,
  $\consetone\models\consettwo$ iff $\consetone\models p\leq q$ for every
  constraint $p\leq q$ in $\consettwo$.
\item
  For each constraint set $\consetone$, we define an order $\lepoly_\consetone$ on 
  resource polynomials by imposing $p\lepoly_\consetone q$ iff 
  $\consetone\models p\leq q$.
\end{enumerate}
\end{defi}

\noindent Resource polynomials are ordered extensionally: $p\leq q$ holds if $p$ is smaller
than $q$ in the standard model of arithmetic. This definition is different from
the one from~\cite{Girard92} which is weaker but syntactical, defining $p$ to be
smaller or equal to $q$ if and only if $q-p$ is itself a resource polynomial. This choice
is motivated by the necessity of reasoning about resource polynomials under some assumptions
in a constraint set. On the other hand, it has some consequences for the decidability 
of type checking, discussed in Section~\ref{sect:typechecking}.

The following is a useful technical result about constraints: 
\begin{lem}\label{lemma:monotoneconst}
If $x_1,\ldots,x_n$ occur positively (negatively, respectively) in $\consetone$,
$p_i\lepoly_{\consettwo}q_i$ for every $i$ and $\multi{x}\notin\FV{\consettwo}$, then 
$\consetone\{\multi{p}/\multi{x}\}\cup\consettwo\models\consetone\{\multi{q}/\multi{x}\}$
($\consetone\{\multi{q}/\multi{x}\}\cup\consettwo\models\consetone\{\multi{p}/\multi{x}\}$,
respectively).
\end{lem}
\proof
Take any constraint $r\leq t$ in $\consetone$ and suppose  
$x_1,\ldots,x_n$ occur positively in $\consetone$.
Then $x_1,\ldots,x_n$ can occur in $t$ but they \emph{cannot}
occur in $r$. So:
\begin{eqnarray*}
(r\leq t)\{\multi{p}/\multi{x}\}&=&r\leq (t\{\multi{p}/\multi{x}\})\\
(r\leq t)\{\multi{q}/\multi{x}\}&=&r\leq (t\{\multi{q}/\multi{x}\})
\end{eqnarray*}
Now, since $p_i\lepoly_{\consettwo}q_i$ for every $i$,
$t\{\multi{p}/\multi{x}\}\lepoly_{\consettwo}t\{\multi{q}/\multi{x}\}$.
As a consequence,
$\consetone\{\multi{p}/\multi{x}\}\cup\consettwo\models\consetone\{\multi{q}/\multi{x}\}$.
Analogously if $x_1,\ldots,x_n$ occur only negatively in $\consetone$.\qed

\subsection{Formulas}
Resource polynomials, constraints and constraint sets are the essential
ingredients in the definition of \QBAL\ formulas:
\begin{defi}\label{defn:formulas}
  Formulas of \QBAL\ are defined as follows:
  \begin{eqnarray*}
  A::=&\alpha(p_1,\ldots,p_n)\spb A\otimes A\spb A\multimap A\spb \forall\alpha.A\spb !_{x< p}A\spb \\
  &\forall(x_1,\ldots,x_n):\consetone.A\spb\exists(x_1,\ldots,x_n):\consetone.A
  \end{eqnarray*}
  where $x\notin\FV{p}$, $\alpha$ ranges over a countable class of atoms (each with an arity 
  $n$). We will restrict ourselves to \emph{bounded} first order quantification. In other words,
  whenever we write 
  $
  \forall(x_1,\ldots,x_n):\consetone.A
  $
  or
  $
  \exists(x_1,\ldots,x_n):\consetone.A
  $
  we implicitly assume that for every $i$ there is a resource polynomial
  $p_i$ not containing the variables $x_1,\ldots,x_n$ such that 
  $\consetone\models\{x_1\leq p_1,\ldots,x_n\leq p_n\}$. 
\end{defi}
Checking the boundedness condition on formulas is undecidable in general (see Section~\ref{sect:typechecking} for further
discussion). The notions of a free atom or a free resource variable in a formula are defined as usual, keeping
in mind that $\forall\overline{x}$, $\exists\overline{x}$ and $!_{x<p}$ act as binders for resource variables, while 
$\forall\alpha$ acts as a binder for atoms.

Notice that resource polynomials and the variables in them can occur inside constraints, constraint
sets and formulas. The following definition becomes natural:
\begin{defi}[Positive and Negative Occurrences]
  The definition of a positive (or negative) free occurrence of a variable 
  in a formula $A$ proceeds by induction on $A$:
  \begin{enumerate}[$\bullet$]
  \item
    All the variables in $\FV{p_1}\cup\ldots\cup\FV{p_n}$ occur positively 
    in $\alpha(p_1,\ldots,p_n)$.
  \item Polarities are propagated through compound formulas by
    noting that $\limp$ is negative in the first slot, $!_{x<p}$
    is negative in $p$ and $\forall \multi{x}{:}\consetone$ is
    negative in $\consetone$. All other slots are positive. 
    We omit the detailed definition. 
  \end{enumerate}
\end{defi}
For example, the first occurrence of  $x$ in 
$\forall y{:}\{y{\leq}x\}.\beta(z)\linear\alpha(x,y)$ is negative while the second one is
positive; the only occurrence of $z$ is negative; all occurrences of $y$ are bound.

When resource variables occur positively in a formula, one can substitute the formula
for an atom in another formula:
\begin{defi}[Substitution]
  Let $B$ be a formula where the free variables $x_1,\ldots,x_n$ occur only
  positively. Then $A\{B/\alpha(x_1,\ldots,x_n)\}$ denotes the formula
  obtained by replacing every free occurrence of $\alpha(p_1,\ldots,p_n)$ with
  $B\{p_1/x_1,\ldots,p_n/x_n\}$ inside $A$.
\end{defi}
As an example , if $A$ is $\alpha(p)\multimap\alpha(p+1)$ and $B$ is
$\exists x:\{x\leq y\}.\beta(y)$, then $A\{B/\alpha(y)\}$ is
$$
(\exists x:\{x\leq p\}.\beta(p))\multimap(\exists x:\{x\leq p+1\}.\beta(p+1)).
$$

Formulas can be compared with respect to assumptions in the form of a constraint
set:
\begin{defi}[Ordering Formulas]
  The order $\lepoly_\consetone$ on resource polynomials can be extended to an order on formulas
  as follows:
  \begin{eqnarray*}
    \alpha(p_1,\ldots,p_n)&\leform_\consetone&\alpha(q_1,\ldots q_n)\mathiff\forall i.p_i\lepoly_\consetone q_i\\
    A\otimes B&\leform_\consetone& C\otimes D\mathiff A\leform_\consetone C\wedge B\leform_\consetone D\\
    A\multimap B&\leform_\consetone& C\multimap D\mathiff C\leform_\consetone A\wedge B\leform_\consetone D\\
    \forall\alpha.A&\leform_\consetone&\forall\alpha.B\mathiff A\leform_\consetone B\\
    !_{x<p}A&\leform_\consetone&!_{x<q}B\mathiff q\lepoly_\consetone p\wedge x\notin\FV{\consetone}\wedge A\leform_{\consetone\cup\{x<q\}} B\\
    \forall\multi{x}:\consettwo.A&\leform_\consetone&\forall\multi{x}:\consetthree.B\mathiff
      \consetone\cup\consetthree\models\consettwo\wedge\multi{x}\notin\FV{\consetone}\wedge A\lepoly_{\consetone\cup\consetthree} B\\
    \exists\multi{x}:\consettwo.A&\leform_\consetone&\exists\multi{x}:\consetthree.B\mathiff
      \consetone\cup\consettwo\models\consetthree\wedge\multi{x}\notin\FV{\consetone}\wedge A\lepoly_{\consetone\cup\consettwo} B
  \end{eqnarray*}
\end{defi}
Please observe that if $A\lepoly_\consetone B$, then $A$ and $B$ have the same ``logical skeleton'' and only
differ in the corresponding resource polynomials and constraint sets. Resource polynomials in positive position 
are smaller in $A$, while those in negative position are smaller in $B$. Moreover, constraint sets in positive position
are stronger in $A$, while those in negative position are stronger in $B$. Consider, as an example, the constraint
set $\consetone=\{x\leq y+1\}$. It is easy to check that
$$
\forall z:\{z\leq y+2\}.\alpha(y+1)\multimap\alpha(z)\lepoly_\consetone\forall z:\{z\leq x\}.\alpha(z)\multimap\alpha(y+2).
$$
Indeed $\consetone\cup\{z\leq x\}\models\{z\leq y+1\}\models\{z\leq y+2\}=\consettwo$, $z\lepoly_{\consetone\cup\consettwo} y+2$
and $z\lepoly_{\consetone\cup\consettwo} y+1$. Intuitively, $\lepoly_\consetone$ can be seen as a subtyping relation
such that subtypes of a formula $A$ are those formulas which are ``smaller'' than $A$ whenever the constraints
in $\consetone$ hold. This is in accordance to, e.g.,
the way $\lepoly_\consetone$ is defined on implicational formulas, which is very reminiscent of the usual rule defining
subtyping for arrow types.

The order relations $\lepoly_\consetone$ satisfy some basic properties:
\begin{lem}[Strengthening and Transitivity]\label{lemma:leform}
If $A\lepoly_\consetone B$ and $\consettwo\models\consetone$, then $A\lepoly_\consettwo B$. Moreover,
if $A\lepoly_\consetone B$ and $B\lepoly_\consetone C$, then $A\lepoly_\consetone C$.
\end{lem}
\proof
Strengthening can be proved by an induction on $A$. Some cases:
\begin{enumerate}[$\bullet$]
\item
  If $!_{x<p}A\leform_\consetone!_{x<q}B$ and $\consettwo\models\consetone$, then
  $q\lepoly_\consetone p$, $x\notin\FV{\consetone}$, and
  $A\leform_{\consetone\cup\{x<q\}} B$. By induction hypothesis, 
  $A\leform_{\consettwo\cup\{x<q\}} B$. We can assume that $x\notin\FV{\consettwo}$. Finally,
  $q\lepoly_\consettwo p$. The thesis easily follows.
\item
  If $\forall\multi{x}:\consettwo.A\leform_\consetone\forall\multi{x}:\consetthree.B$
  and $\consetfour\models\consetone$, then $\consetone\cup\consetthree\models\consettwo$, 
  $\multi{x}\notin\FV{\consetone}$ and $A\lepoly_{\consetone\cup\consetthree} B$.
  Again, we can assume that $\multi{x}\notin\FV{\consetone}$. Since
  $\consetfour\cup\consetthree\models\consetone\cup\consetthree$, we can apply
  the induction hypothesis, obtaining $A\lepoly_{\consetfour\cup\consetthree} B$,
  from which the thesis easily follows.
\end{enumerate}
Transitivity can be handled itself by induction on the structure of $A$. Some cases:
\begin{enumerate}[$\bullet$]
\item
  If $!_{x<p}A\leform_\consetone!_{x<q}B$ and $!_{x<q}B\leform_\consetone!_{x<r}C$, then
  $q\lepoly_\consetone p$, $r\lepoly_\consetone q$, $x\notin\FV{\consetone}$,
  $A\leform_{\consetone\cup\{x<q\}} B$ and $B\leform_{\consetone\cup\{x<r\}}C$.
  From $A\leform_{\consetone\cup\{x<q\}}B$ and $r\lepoly_\consetone q$, it follows 
  by strengthening that $A\leform_{\consetone\cup\{x<r\}}B$. By the induction
  hypothesis, $A\leform_{\consetone\cup\{x<r\}}C$. The thesis follows, once
  we observe that $r\lepoly_\consetone p$ by transitivity.
\item
  If $\forall\multi{x}:\consettwo.A\leform_\consetone\forall\multi{x}:\consetthree.B$
  and $\forall\multi{x}:\consetthree.B\leform_\consetone\forall\multi{x}:\consetfour.C$,
  then $\consetone\cup\consetthree\models\consettwo$, $\consetone\cup\consetfour\models\consetthree$,
  $\multi{x}\notin\FV{\consetone}$, $A\lepoly_{\consetone\cup\consetthree} B$ and
  $B\lepoly_{\consetone\cup\consetfour} C$. Since $\consetone\cup\consetfour\models\consetone\cup\consetthree$,
  $A\lepoly_{\consetone\cup\consetfour} B$ by strengthening. By the induction hypothesis,
  $A\lepoly_{\consetone\cup\consetfour} C$. The thesis follows observing that $\consetone\cup\consetfour\models\consettwo$.
\item
  If $\exists\multi{x}:\consettwo.A\leform_\consetone\exists\multi{x}:\consetthree.B$
  and $\exists\multi{x}:\consetthree.B\leform_\consetone\exists\multi{x}:\consetfour.C$,
  then $\consetone\cup\consettwo\models\consetthree$, $\consetone\cup\consetthree\models\consetfour$,
  $\multi{x}\notin\FV{\consetone}$, $A\lepoly_{\consetone\cup\consettwo} B$ and
  $B\lepoly_{\consetone\cup\consetthree} C$. Since $\consetone\cup\consettwo\models\consetone\cup\consetthree$,
  $B\lepoly_{\consetone\cup\consettwo} C$ by strengthening. By the induction hypothesis,
  $A\lepoly_{\consetone\cup\consettwo} C$. The thesis follows observing that $\consetone\cup\consettwo\models\consetfour$.
\end{enumerate}
This concludes the proof.\qed

Some other auxiliary results about the relations $\leform_\consetone$ will be useful in the following.
We give them here. First of all, we can perform substitution of resource polynomials for resource
variables in formulas being sure that the underlying order is preserved:
\begin{lem}[Monotonicity, I]\label{lemma:lepolysubst}
If $A$ is a formula where the variables $x_1,\ldots,x_n$ occur only
positively (negatively, respectively), 
$p_i\lepoly_{\consetone}q_i$ for every $i$ and $\multi{x}\notin\FV{\consetone}$, then 
$A\{\multi{p}/\multi{x}\}\lepoly_{\consetone}A\{\multi{q}/\multi{x}\}$
($A\{\multi{q}/\multi{x}\}\lepoly_{\consetone}A\{\multi{p}/\multi{x}\}$,
respectively).
\end{lem}
\proof
By induction on $A$. Let's just check the most interesting cases:
\begin{enumerate}[$\bullet$]
\item
  If $A=\exists\multi{y}:\consettwo.B$, then the variables in $\multi{y}$ can be assumed to be
  distinct from $x_1\ldots,x_n$. Moreover:
  \begin{eqnarray*}
    A\{\multi{p}/\multi{x}\}&=&\exists\multi{y}:\consettwo\{\multi{p}/\multi{x}\}.B\{\multi{p}/\multi{x}\}\\
    A\{\multi{q}/\multi{x}\}&=&\exists\multi{y}:\consettwo\{\multi{q}/\multi{x}\}.B\{\multi{q}/\multi{x}\}
  \end{eqnarray*}
  Now, suppose that $x_1,\ldots,x_n$ occur only
  positively in $A$. Then, by induction hypothesis, $B\{\multi{p}/\multi{x}\}\lepoly_{\consetone}B\{\multi{q}/\multi{x}\}$.
  Moreover, by Lemma~\ref{lemma:monotoneconst},
  $\consettwo\{\multi{p}/\multi{x}\}\cup\consetone\models\consettwo\{\multi{q}/\multi{x}\}$. By definition,
  this implies $A\{\multi{p}/\multi{x}\}\lepoly_{\consetone}A\{\multi{q}/\multi{x}\}$.
  Similarly if $x_1,\ldots,x_n$ occur only negatively in $A$.
\item
  If $A=\forall\multi{y}:\consettwo.B$, then we can proceed exactly in the same way.
\end{enumerate}
This concludes the proof.\qed

On the other hand, the same polynomial can be substituted in formulas, again preserving
the underlying order:
\begin{lem}[Monotonicity, II]\label{lemma:leformsubst}
If $A\leform_{\consetone} B$ then
$A\{\multi{p}/\multi{x}\}\leform_{\consetone\{\multi{p}/\multi{x}\}} B\{\multi{p}/\multi{x}\}$.
\end{lem}
\proof
By induction on $A$. Some interesting cases:
\begin{enumerate}[$\bullet$]
\item
   If $A=!_{x<q}C$, then $B=!_{x<r}D$, $r\lepoly_{\consetone}q$, $x\notin\FV{\consetone}$,
   $C\leform_{\consetone\cup\{x<r\}}D$ and $x$ can be assumed not to appear among the variables
   in $\multi{x}$ nor the ones in $\multi{p}$. Then:
   \begin{eqnarray*}
     A\{\multi{p}/\multi{x}\}&=!_{x<q\{\multi{p}/\multi{x}\}}C\{\multi{p}/\multi{x}\}\\
     B\{\multi{p}/\multi{x}\}&=!_{x<r\{\multi{p}/\multi{x}\}}D\{\multi{p}/\multi{x}\}
   \end{eqnarray*}
   By induction hypothesis, $C\{\multi{p}/\multi{x}\}\leform_{\consetone\{\multi{p}/\multi{x}\}\cup\{x<r\{\multi{p}/\multi{x}\}\}}D\{\multi{p}/\multi{x}\}$.
   Moreover, $r\{\multi{p}/\multi{x}\}\lepoly_{\consetone\{\multi{p}/\multi{x}\}}q\{\multi{p}/\multi{x}\}$.
   The thesis follows.
\item
  If $A=\exists\multi{y}:\consettwo.C$, then $B=\exists\multi{y}:\consetthree.D$,
  $\consetone\cup\consettwo\models\consetthree$, $\multi{y}\notin\FV{\consetone}$,
  $C\leform_{\consetone\cup\consetthree}D$ and, again, variables in $\multi{y}$ can be assumed
  not to appear among the variables in $\multi{x}$ nor in the ones in $\multi{p}$. Then:
  \begin{eqnarray*}
     A\{\multi{p}/\multi{x}\}&=\exists\multi{y}:\consettwo\{\multi{p}/\multi{x}\}.C\{\multi{p}/\multi{x}\}\\
     B\{\multi{p}/\multi{x}\}&=\exists\multi{y}:\consetthree\{\multi{p}/\multi{x}\}.D\{\multi{p}/\multi{x}\}
  \end{eqnarray*}
  Clearly, $\consetone\{\multi{p}/\multi{x}\}\cup\consettwo\{\multi{p}/\multi{x}\}\models\consetthree\{\multi{p}/\multi{x}\}$.
  By the assumptions above, $\multi{y}\notin\FV{\consetone\{\multi{p}/\multi{x}\}}$.
  By the induction hypothesis, 
  $C\{\multi{p}/\multi{x}\}\leform_{\consetone\{\multi{p}/\multi{x}\}\cup\consetthree\{\multi{p}/\multi{x}\}}D\{\multi{p}/\multi{x}\}$.
  The thesis follows.
\end{enumerate}
This concludes the proof.\qed

Finally, substitution of formulas for \emph{atoms} preserves itself the order $\leform_\consetone$:
\begin{lem}[Monotonicity, III]\label{lemma:leformsubstII}
If $C$ is a formula where the free variables $x_1,\ldots,x_n$ occur only
positively, $\alpha$ is an atom with arity $n$ 
and $A\leform_{\consetone} B$, then $A\{C/\alpha\}\leform_{\consetone} B\{C/\alpha\}$.
\end{lem}
\proof
By induction on $A$. Some cases:
\begin{enumerate}[$\bullet$]
\item
  If $A=\alpha(\multi{p})$, then $B=\alpha(\multi{q})$ where
  $p_i\lepoly_\consetone q_i$, $A\{C/\alpha\}=C\{\multi{p}/\multi{x}\}$
  and $B\{C/\alpha\}=C\{\multi{q}/\multi{x}\}$. The thesis easily follows
  from Lemma~\ref{lemma:lepolysubst}.
\item
  If $A=\forall\beta.D$, then $B=\forall\beta.E$ and
  we can assume that $\beta\neq\alpha$ and
  that $\beta$ does not appear free in $C$. Then 
  $A\{C/\alpha\}=\forall\beta.(D\{C/\alpha\})$ and
  $B\{C/\alpha\}=\forall\beta.(E\{C/\alpha\})$. From
  $A\leform_\consetone B$, it follows that $D\leform_\consetone E$ and,
  by induction hypothesis, that $D\{C/\alpha\}\leform_\consetone E\{C/\alpha\}$.
  The thesis easily follows.
\end{enumerate}
This concludes the proof.\qed

\subsection{Rules}
A \QBAL\ \emph{judgement} is an expression in the form $\judgebll{\consetone}{\Gamma}{A}$,
where $\consetone$ is a constraint set, $\Gamma$ is a multiset of formulas and $A$
is a formula. The meaning of such an expression is the following: $A$ is a consequence of
the formulas in $\Gamma$, provided the constraints in $\consetone$ hold.

\emph{Rules of inference} for \QBAL\ are in Figure~\ref{figure:qbal}.
\begin{figure*}[htbp]
\begin{center}
\fbox{
\begin{minipage}{.97\textwidth}
\centering\textbf{Axiom and Cut}
$$
\begin{array}{ccc}
\infer[A]{\judgebll{\consetone}{A}{B}}{A\leform_\consetone B} 
&\hspace{5mm}&
\infer[U]
{\judgebll{\consetone}{\Gamma,\Delta}{B}} 
{\judgebll{\consetone}{\Gamma}{A} & \judgebll{\consetone}{\Delta,A}{B}}
\end{array}
$$
\centering\textbf{Structural Rules}
$$
\begin{array}{ccc}
\infer[W]
{\judgebll{\consetone}{\Gamma,A}{B}}
{\judgebll{\consetone}{\Gamma}{B}}
&\hspace{3mm}&
\infer[X]
{\judgebll{\consetone}{\Gamma,!_{x<r}A}{B}}
{\judgebll{\consetone}{\Gamma,!_{x<p}A,!_{y< q}A\{p+y/x\}}{B} & p+q\sqsubseteq_\consetone r}
\end{array}
$$
\centering\textbf{Multiplicative Logical Rules}
$$
\begin{array}{ccc}
\infer[R_\linear]
{\judgebll{\consetone}{\Gamma}{A\linear B}}
{\judgebll{\consetone}{\Gamma,A}{B}} 
&\hspace{4mm}&
\infer[L_\linear]
{\judgebll{\consetone}{\Gamma,\Delta,A\linear B}{C}} 
{\judgebll{\consetone}{\Gamma}{A} & \judgebll{\consetone}{\Delta,B}{C}}
\end{array}
$$
$$
\begin{array}{ccc}
\infer[R_\otimes]
{\judgebll{\consetone}{\Gamma,\Delta}{A\otimes B}}
{\judgebll{\consetone}{\Gamma}{A}\ & \judgebll{\consetone}{\Delta}{B}} 
&\hspace{4mm}&
\infer[L_\otimes]
{\judgebll{\consetone}{\Gamma,A\otimes B}{C}} 
{\judgebll{\consetone}{\Gamma,A,B}{C}}
\end{array}
$$
\centering\textbf{Exponential Rules}
$$
\infer[P_{!}]
{\judgebll{\consettwo}{!_{x<q_1}A_1,\ldots,!_{x<q_n}A_n}{!_{x<p}B}}
{\judgebll{\consetone}{A_1,\ldots,A_n}{B} & \consettwo\cup\{x<p\}\models\consetone & x\notin\FV{\consettwo} & p\lepoly_\consettwo q_i}
$$
$$
\infer[D_{!}]
{\judgebll{\consetone}{!_{x<p}A,\Gamma}{B}}
{\judgebll{\consetone}{A\{1/x\},\Gamma}{B} & 1\lepoly_\consetone p}
$$
\vspace{1mm}
$$
\infer[N_{!}]
{\judgebll{\consetone}{!_{x<r}A,\Gamma}{B}}
{\judgebll{\consetone}{!_{y<p}!_{z<q\{y/w\}}A\{(z+\sum_{w<y}q)/x\},\Gamma}{B} & \sum_{w<p}q\lepoly_\consetone r}
$$
\centering\textbf{Second Order Rules}
$$
\begin{array}{ccc}
\infer[R_{\forall\alpha}]
{\judgebll{\consetone}{\Gamma}{\forall\alpha.A}}
{\judgebll{\consetone}{\Gamma}{A} & \alpha\not\in\mathit{FV}(\Gamma)}
&\hspace{2mm}&
\infer[L_{\forall\alpha}]
{\judgebll{\consetone}{\Gamma,\forall\alpha.A}{C}}
{\judgebll{\consetone}{\Gamma,A\{B/\alpha(x_1,\ldots,x_n)\}}{C}}
\end{array}
$$
\centering\textbf{First Order Rules}
$$
\begin{array}{ccc}
\infer[R_{\forall x}]
{\judgebll{\consetone}{\Gamma}{\forall\multi{x}:\consettwo.A}}
{\judgebll{\consetone\cup\consettwo}{\Gamma}{A} & \multi{x}\not\in\FV{\Gamma}\cup\FV{\consetone}}
&&
\infer[L_{\forall x}]
{\judgebll{\consetone}{\Gamma,\forall\multi{x}:\consettwo.A}{C}}
{\judgebll{\consetone}{\Gamma,A\{\multi{p}/\multi{x}\}}{C} & \consetone\models\consettwo\{\multi{p}/\multi{x}\}}
\end{array}
$$
$$
\begin{array}{ccc}
\infer[\!R_{\exists x}]
{\judgebll{\consetone}{\Gamma}{\exists\multi{x}:\consettwo.A}}
{\judgebll{\consetone}{\Gamma}{A\{\multi{p}/\multi{x}\}} & \consetone\models\consettwo\{\multi{p}/\multi{x}\}}
&\!\!\!&
\infer[\!L_{\exists x}]
{\judgebll{\consetone}{\Gamma,\exists\multi{x}:\consettwo.A}{C}}
{\judgebll{\consetone\cup\consettwo}{\Gamma,A}{C} &  \multi{x}\not\in\FV{\Gamma}\cup\FV{C}\cup\FV{\consetone}}
\end{array}
$$
\end{minipage}}
\caption{A sequent calculus for $\QBAL$}
\label{figure:qbal}
\end{center}
\end{figure*}
All rules except first order ones are the natural generalizations of \BLL\ rules. In particular, observe that
the only rules modifying the underlying constraint set(s) are $P_!$, $R_{\forall x}$, 
$L_{\forall x}$, $R_{\exists x}$ and $L_{\exists x}$. The multiplicative connectives are governed by the usual
rules from intuitionistic linear logic. The modality $!$, on the other hand, is a functor governed by the following
axioms, which come from \BLL:
\begin{eqnarray*}
!_{x<1}A&\multimap A\{1/x\};\\
!_{x<p+q}A&\multimap!_{x<p}A\otimes!_{y<q}A\{p+y/x\};\\
!_{x<\sum_{w<p}q}A&\multimap !_{y<p}!_{z<q\{y/w\}}A\{(z+\sum_{w<y}q)/x\}.
\end{eqnarray*}
Moreover, given a constraint set $\consetone$, it holds that $!_{p}A\multimap !_{q}A$ 
whenever $q\lepoly_\consetone p$. Weakening holds for every formula, contrary to what happens in \BLL;
as is usual in systems derived from linear logic, this does not break good quantitative properties like 
polynomial time soundness. Rules $W$, $X$, $P_!$, $D_!$ and $N_!$ capture the just described behaviour. Observe
how rule $P_!$ allows to take advantage of the inequality $x<q$ in the premise.

First order quantification on resource variables, on the other hand, is governed by the four inference
rules $R_{\forall x}$, $L_{\forall x}$, $R_{\exists x}$ and $L_{\exists x}$. Let us consider, as an example,
rule $R_{\forall x}$, which can be read as follows: if $A$ can be inferred from $\Gamma$ provided
$\consetone\cup\consettwo$ hold \emph{and} the variables $\multi{x}$ do not appear in $\Gamma$ nor
in $\consetone$, then $A$ holds for every value of $\multi{x}$ satisfying $\consetone$.

Notice that \BLL\ can be embedded into \QBAL: for every \BLL\ proof $\pi:\Gamma\vdash A$, 
there is a \QBAL\ proof $\trlqbllpr{\pi}:\judgebll{\emptyset}{\Gamma}{A}$: this
can be proved by an easy induction on $\pi$.

If $\pi$ is a proof of \QBAL, then $|\pi|$ is the number of rule instances in $\pi$.
\subsection{On Decidability of Proof Checking}\label{sect:typechecking}
The problem of checking the correctness of a proof is undecidable in \QBAL, since the
correctness of a formula is itself an undecidable problem: remember that a
formula like $\exists x:\consetone.A$ is correct only if an inequality $x\leq p$ can
be deduced from $\consetone$ for \emph{some} polynomial $p$.
Moreover, the relation $\models$ between constraint sets is undecidable.
This is in contrast to what happens in more implicit ICC systems or in \BLL\ itself,
where conditional equality $\lepoly_\consetone$ is replaced by unconditional inequality
and $p\leq q$ iff $q-p$ is a resource polynomial.

We do not see undecidability of proof checking 
as a fundamental problem of \QBAL\, for at least two reasons:
\begin{enumerate}[$\bullet$]
\item
  On the one hand, the main role of \QBAL\ is the one of a metasystem in which
  to prove quantitative properties of other systems. As a consequence, it is
  crucial to keep the system as expressive as possible.
\item
  On the other hand, simple, decidable fragments of \QBAL\ can possibly be
  built by considering decidable, although necessarily incomplete, formal
  systems for assertions in the form $\consetone\vdash p\leq q$ (or,
  more generally, $\consetone\vdash\consettwo$) and by imposing that
  bounds on quantified variables are given explicitly when forming existential
  or universal formulas. This, however, is a topic outside the scope of this paper, which
  we leave for future work.
\end{enumerate}
The way we define \QBAL\ is, in other words, similar to the one Xi adopts
when he introduces Dependent \ML~\cite{Xi07}.
\subsection{\QBAL\ and Second Order Logic}\label{sect:qbllsol}
Second order intuitionistic logic can be presented as a context-independent sequent
calculus with explicit structural rules~\cite{Troelstra96}, \Gtwoi. Rules of \Gtwoi\ are
in Figure~\ref{figure:gtwoi}.
There is a forgetful map $\forget{\cdot}$ from the space of \QBAL\ proofs to the space
of \Gtwoi\ proofs. In particular $\linear$ corresponds to $\rightarrow$ and
$\otimes$ corresponds to $\wedge$. Essentially, $\forget{\pi}$ has the same structure
as $\pi$, except for exponential and first order rules, which have no formal correspondence
in \Gtwoi. From our point of view, if $\forget{\pi}=\forget{\rho}$, then $\pi$ and $\rho$
correspond to the same \emph{program}, i.e. \QBAL\ can be seen as a proper decoration of
second order logic proofs with additional information which is not necessary to perform
the underlying computation. 
\begin{figure*}[htbp]
\begin{center}
\fbox{
\begin{minipage}{.97\textwidth}
\centering\textbf{Axiom, Cut and Structural Rules}
$$
\begin{array}{ccccccc}
\infer[A]{A\vdash A}{} 
&&
\infer[U]
{\Gamma,\Delta\vdash B} 
{\Gamma\vdash A & \Delta,A\vdash B}
&&
\infer[W]
{\Gamma,A\vdash B}
{\Gamma\vdash B}
&&
\infer[X]
{\Gamma,A\vdash B}
{\Gamma,A,A\vdash B}
\end{array}
$$
\centering\textbf{Logical Rules}
$$
\begin{array}{ccccccc}
\infer[R_\rightarrow]
{\Gamma\vdash A\rightarrow B}
{\Gamma,A\vdash B} 
&\!\!&
\infer[L_\rightarrow]
{\Gamma,\Delta,A\rightarrow B\vdash C} 
{\Gamma\vdash A & \Delta,B\vdash C}
&\!\!&
\infer[R_\times]
{\Gamma,\Delta\vdash A\times B}
{\Gamma\vdash A\ & \Delta\vdash B} 
&\!\!&
\infer[L_\times]
{\Gamma,A\times B\vdash C} 
{\Gamma,A,B\vdash C}
\end{array}
$$
\centering\textbf{Second Order Rules}
$$
\begin{array}{ccc}
\infer[R_{\forall}]
{\Gamma\vdash \forall\alpha.A}
{\Gamma\vdash A & \alpha\not\in\mathit{FV}(\Gamma)}
&&
\infer[L_{\forall}]
{\Gamma,\forall\alpha.A\vdash C}
{\Gamma,A\{B/\alpha\}\vdash C}
\end{array}
$$
\end{minipage}}
\caption{A sequent calculus for $\Gtwoi$}
\label{figure:gtwoi}
\end{center}
\end{figure*}
\subsection{Properties}\label{sect:properties}
\QBAL\ inherits some nice properties from \BLL. In particular, proofs can be manipulated
in a uniform way by altering their conclusion without changing their structure, i.e., without
changing the underlying second order logic proof.

First of all, a useful transformation is the strengthening of the
underlying constraint set $\consetone$:
\begin{prop}[Strengthening]\label{prop:strengthcon}
If $\pi:\judgebll{\consetone}{\Gamma}{B}$ is a proof
and $\consettwo\models\consetone$, then there is a proof 
$\rho:\judgebll{\consettwo}{\Gamma}{B}$ such that
$\forget{\rho}=\forget{\pi}$ and $|\rho|\leq |\pi|$ .
\end{prop}
\proof
An easy induction on $\pi$. As an example, if $\pi$ consists of an instance of rule $A$, then
the thesis follows from Lemma~\ref{lemma:leform}. As another example, take a proof 
$\pi:\judgebll{\consetone}{\Gamma}{\forall\multi{x}:\consettwo.A}$ obtained
from $\rho:\judgebll{\consetone\cup\consetthree}{\Gamma}{A}$ applying rule $R_{\forall x}$.
We can assume without losing generality that $\multi{x}\notin\FV{\consettwo}$.
From $\consettwo\models\consetone$, it follows that 
$\consettwo\cup\consetthree\models\consetone\cup\consetthree$, from which the thesis easily follows.\qed

\QBAL\ is monotone with respect to the relation $\lepoly_\consetone$ on
formulas. 
\begin{prop}[Monotonicity]\label{prop:monotonicity}
If $\pi:\judgebll{\consetone}{A_1,\ldots,A_n}{B}$, $B\lepoly_\consetone D$ and 
$C_i\lepoly_\consetone A_i$ for
every $1\leq i\leq n$, then there is $\rho:\judgebll{\consetone}{C_1,\ldots,C_n}{D}$ such
that $\forget{\rho}=\forget{\pi}$.
\end{prop}
\proof
By induction on $|\pi|$.
Some interesting cases are the following ones:
\begin{enumerate}[$\bullet$]
\item
  Suppose that $\pi$ is simply
  $$
  \infer[]{\judgebll{\consetone}{A}{B}}{A\leform_\consetone B}
  $$
  and that $C\lepoly_\consetone A$ and $B\lepoly_\consetone D$. Then, by
  transitivity of $\leform_\consetone$ (see Lemma~\ref{lemma:leform}), $C\lepoly_\consetone D$ and $\rho$ is 
  $$
  \infer[]{\judgebll{\consetone}{C}{D}}{C\leform_\consetone D}.
  $$
\item
  Suppose that $\pi$ is
  $$
  \infer[]
  {\judgebll{\consettwo}{!_{x<q_1}A_1,\ldots,!_{x<q_n}A_n}{!_{x<p}B}}
  {\judgebll{\consetone}{A_1,\ldots,A_n}{B} & \consettwo\cup\{x<p\}\models\consetone & x\notin\FV{\consettwo} & p\lepoly_\consettwo q_i}
  $$
  and that $!_{x<r_i}C_i\lepoly_\consettwo !_{x<q_i}A_i$ (for every $i$) and
  $!_{x<p}B\lepoly_\consettwo !_{x<s} D$. By definition, $q_i\lepoly_\consettwo r_i$ for every $i$
  and $s\lepoly_\consettwo p$. By the side condition to the premise of $\pi$ (and by transitivity of
  $\lepoly_\consettwo$), we obtain $s\lepoly_\consettwo r_i$ for every $i$.
  Moreover, we have $C_i\lepoly_{\consettwo\cup\{x<q_i\}} A_i$ for every $i$
  and $B\lepoly_{\consettwo\cup\{x<s\}} D$. This implies
  $C_i\lepoly_{\consettwo\cup\{x<s\}} A_i$ for every $i$, because
  $$
  \consettwo\cup\{x<s\}\models\consettwo\cup\{x<p\}\models\consettwo\cup\{x<q_i\}
  $$
  and by Lemma~\ref{lemma:leform}.
  Now, since $\consettwo\cup\{x<s\}\models\consettwo\cup\{x<p\}\models\consetone$,
  we can obtain, by Proposition~\ref{prop:strengthcon}, a proof
  $\sigma$ of $\judgebll{\consettwo\cup\{x<s\}}{A_1,\ldots,A_n}{B}$ such that $|\sigma|<|\pi|$.
  Then, we can easily apply the induction hypothesis on $\sigma$ and conclude.
\end{enumerate}
This concludes the proof.\qed

Another useful transformation on proofs is the substitution of resource polynomials for free variables.
\begin{prop}[Substitution for Variables]\label{prop:substpol}
If $\pi:\judgebll{\consetone}{A_1,\ldots,A_n}{B}$ is a proof and $p_1,\ldots,p_n$ are resource polynomials, 
then there is a proof 
\[\pi\{\multi{p}/\multi{x}\}:\judgebll{\consetone\{\multi{p}/\multi{x}\}}
{A_1\{\multi{p}/\multi{x}\},\ldots,A_n\{\multi{p}/\multi{x}\}}{B\{\multi{p}/\multi{x}\}}
\]
 such that $\forget{\pi\{\multi{p}/\multi{x}\}}=\forget{\pi}$.
\end{prop}
\proof
By induction on $\pi$. An interesting case is the following one:
\begin{enumerate}[$\bullet$]
\item
  Suppose that $\pi$ is simply
  $$
  \infer[]{\judgebll{\consetone}{A}{B}}{A\leform_\consetone B}.
  $$
  Now, if $A\leform_{\consetone} B$ then
  $A\{\multi{p}/\multi{x}\}\leform_{\consetone\{\multi{p}/\multi{x}\}} B\{\multi{p}/\multi{x}\}$, 
  by Lemma~\ref{lemma:leformsubst}. As a consequence,
  $$
  \infer[]{\judgebll{\consetone\{\multi{p}/\multi{x}\}}{A\{\multi{p}/\multi{x}\}}
    {B\{\multi{p}/\multi{x}\}}}{A\{\multi{p}/\multi{x}\}\leform_{\consetone\{\multi{p}/\multi{x}\}} B\{\multi{p}/\multi{x}\}}.
  $$
\item
  Suppose that $\pi$ is
  $$
  \infer[R_{\exists x}]
  {\judgebll{\consetone}{\Gamma}{\exists\multi{y}:\consettwo.C}}
  {\rho:\judgebll{\consetone}{\Gamma}{C\{\multi{q}/\multi{y}\}} & \consetone\models\consettwo\{\multi{q}/\multi{y}\}}
  $$
  where, without losing generality, $\multi{y}$ can be chosen as fresh variables not appearing among
  the ones in $\multi{x}$, nor in the ones in $\multi{p}$. Applying the induction hypothesis to $\rho$,
  we obtain a proof of 
  $\judgebll{\consetone\{\multi{p}/\multi{x}\}}{\Gamma\{\multi{p}/\multi{x}\}}{C\{\multi{q}/\multi{y}\}\{\multi{p}/\multi{x}\}}$.
  But, by the assumptions above,
  $$
  C\{\multi{q}/\multi{y}\}\{\multi{p}/\multi{x}\}=C\{\multi{p}/\multi{x}\}\{(\multi{q}\{\multi{p}/\multi{x}\})/\multi{y}\}.
  $$
  Analogously, from $\consetone\models\consettwo\{\multi{q}/\multi{y}\}$, it follows that
  $\consetone\{\multi{p}/\multi{x}\}\models\consettwo\{\multi{q}/\multi{y}\}\{\multi{p}/\multi{x}\}$ and
  $$
  \consettwo\{\multi{q}/\multi{y}\}\{\multi{p}/\multi{x}\}=\consettwo\{\multi{p}/\multi{x}\}\{(\multi{q}\{\multi{p}/\multi{x}\})/\multi{y}\}.
  $$
  The thesis follows.
\item
  Suppose that $\pi$ is
  $$
  \infer[L_{\exists x}]
  {\judgebll{\consetone}{\Gamma,\exists\multi{y}:\consettwo.C}{D}}
  {\rho:\judgebll{\consetone\cup\consettwo}{\Gamma,C}{D} &  \multi{y}\not\in\FV{\Gamma}\cup\FV{D}\cup\FV{\consetone}}
  $$
  where, as usual, $\multi{y}$ can be chosen as fresh variables not appearing among
  the ones in $\multi{x}$, nor in $\multi{p}$. Applying the induction hypothesis to $\rho$,
  we obtain a proof of 
  $$
  \judgebll{\consetone\{\multi{p}/\multi{x}\}\cup\consettwo\{\multi{p}/\multi{x}\}}{\Gamma\{\multi{p}/\multi{x}\},C\{\multi{p}/\multi{x}\}}
  {D\{\multi{p}/\multi{x}\}}.
  $$ 
  By the assumptions above, variables in $\multi{y}$ do not appear free
  in $\Gamma\{\multi{p}/\multi{x}\}$ nor in $D\{\multi{p}/\multi{x}\}$ nor in $\consetone\{\multi{p}/\multi{x}\}$.
  The thesis follows.
\end{enumerate}
This concludes the proof.\qed

But formulas themselves can be substituted (for atoms) into a proof:
\begin{prop}[Substitution for Atoms]\label{prop:substform}
If $\pi:\judgebll{\consetone}{A_1,\ldots,A_n}{B}$ is a proof, 
$C$ is a formula where the free variables $x_1,\ldots,x_m$ occur only
positively and $\alpha$ is an atom with arity $m$, then there is a proof 
$\pi\{C/\alpha\}:\judgebll{\consetone}{A_1\{C/\alpha\},\ldots,A_n\{C/\alpha\}}{B\{C/\alpha\}}$ such
that $\forget{\pi\{C/\alpha\}}=\forget{\pi}$.
\end{prop}
\proof
By induction on $\pi$. Some interesting cases are the following ones:
\begin{enumerate}[$\bullet$]
\item
  Suppose, again, that $\pi$ is simply
  $$
  \infer[]{\judgebll{\consetone}{A}{B}}{A\leform_\consetone B}
  $$
  Now, if $A\leform_{\consetone} B$ then
  $A\{C/\alpha\}\leform_{\consetone} B\{C/\alpha\}$
  by Lemma~\ref{lemma:leformsubstII}.
  As a consequence:
  $$
  \infer[]{\judgebll{\consetone}{A\{C/\alpha\}}{B\{C/\alpha\}}}{A\{C/\alpha\}\leform_{\consetone} B\{C/\alpha\}}
  $$
\item
  Suppose that $\pi$ is
  $$
  \infer[R_{\exists x}]
  {\judgebll{\consetone}{\Gamma}{\exists\multi{y}:\consettwo.D}}
  {\rho:\judgebll{\consetone}{\Gamma}{D\{\multi{q}/\multi{y}\}} & \consetone\models\consettwo\{\multi{q}/\multi{y}\}}
  $$
  where, without losing generality, $\multi{y}$ can be chosen as fresh variables not appearing among
  the ones in $\multi{x}$ nor in $C$. Applying the induction hypothesis to $\rho$,
  we obtain a proof of
  $\judgebll{\consetone}{\Gamma\{C/\alpha\}}{D\{\multi{q}/\multi{y}\}\{C/\alpha\}}$.
  But, by the assumptions above,
  $$
  D\{\multi{q}/\multi{y}\}\{C/\alpha\}=D\{C/\alpha\}\{\multi{q}/\multi{y}\}.
  $$
  The thesis follows.
\item
  Suppose that $\pi$ is
  $$
  \infer[L_{\exists x}]
  {\judgebll{\consetone}{\Gamma,\exists\multi{y}:\consettwo.D}{E}}
  {\rho:\judgebll{\consetone\cup\consettwo}{\Gamma,D}{E} &  \multi{y}\not\in\FV{\Gamma}\cup\FV{E}\cup\FV{\consetone}}
  $$
  where, as usual, $\multi{y}$ can be chosen as fresh variables not appearing among
  the ones in $\multi{x}$, nor in $C$. Applying the induction hypothesis to $\rho$,
  we obtain a proof of 
  $$
  \judgebll{\consetone\cup\consettwo}{\Gamma\{C/\alpha\},D\{C/\alpha\}}
  {E\{C/\alpha\}}.
  $$ 
  By the assumptions above, variables in $\multi{y}$ do not appear free
  in $\Gamma\{C/\alpha\}$ nor in $E\{C/\alpha\}$ nor in $\consetone\{C/\alpha\}$: they
  are either in $\Gamma$, $E$, $\consetone$ or in $C$.
  The thesis follows.  
\end{enumerate}
This concludes the proof.\qed

\subsection{Cut-Elimination}
A nice application of the results we have just given is cut-eli\-mi\-na\-tion.
Indeed, the new rules $R_{\forall x}$, $L_{\forall x}$, $R_{\exists x}$ and 
$L_{\exists x}$ do not cause any problem in the cut-elimination process.
For example, the cut
$$
\infer[U]
{
  \judgebll{\consetone}{\Gamma,\Delta}{C}
}
{
  \infer[R_{\forall x}]
  {
    \judgebll{\consetone}{\Gamma}{\forall\multi{x}:\consettwo.A}
  }
  {
    \pi:\judgebll{\consetone\cup\consettwo}{\Gamma}{A} & \multi{x}\notin\FV{\Gamma}\cup\FV{\consetone}
  }
  &
  \infer[L_{\forall x}]
  {
    \judgebll{\consetone}{\Delta,\forall\multi{x}:\consettwo.A}{C}
  }
  {
    \judgebll{\consetone}{\Delta,A\{\multi{p}/\multi{x}\}}{C} & \consetone\models\consettwo\{\multi{p}/\multi{x}\}
  }
}
$$
can be eliminated as follows:
$$
\infer[U]
{
  \judgebll{\consetone}{\Gamma,\Delta}{C}
}
{
  \sigma:\judgebll{\consetone}{\Gamma}{A\{\multi{p}/\multi{x}\}}
  &
  \rho:\judgebll{\consetone}{\Delta,A\{\multi{p}/\multi{x}\}}{C}
}
$$
where $\sigma$ is obtained by applying Proposition~\ref{prop:strengthcon} to
$$
\pi\{\multi{p}/\multi{x}\}:\judgebll{\consetone\cup\consettwo\{\multi{p}/\multi{x}\}}{\Gamma}{A\{\multi{p}/\multi{x}\}},
$$
itself obtained from $\pi$ applying Proposition~\ref{prop:substpol}.
In this paper, we will not study cut-elimination. And polynomial time soundness will be itself proved semantically.
\subsection{Programming in \QBAL}\label{sect:programming}
The Curry-Howard correspondence allows to see \BLL\ and \QBAL\ as programming languages endowed with rich
type systems.
In particular, following the usual impredicative encoding of data into second order intuitionistic logic, natural numbers
can be represented as cut-free proofs of the formula
$$
\Nat_p=\forall\alpha.!_{y<p}(\alpha(y)\linear\alpha(y+1))\linear\alpha(0)\linear\alpha(p).
$$
However, only natural numbers less or equal to $p$ are representable this way. 

This can be generalized to any word algebra.
Given a word algebra $\wa$, we will denote by $\zero_\wa$ the only $0$-ary constructor of
$\wa$ and by $c_\wa^1,\ldots,c_\wa^w$ the $1$-ary constructors of the same algebra. Notice
that these objects can be thought of both as term formers and as ($0$-ary or unary) functions on terms.
Terms of a free algebra $\wa$ of length at most $p$ can be represented as cut-free proofs of the formula
$$
\Wordalg_p=\underbrace{\forall\alpha.!_{y<p}(\alpha(y)\linear\alpha(y+1))\linear\ldots\linear !_{y<p}(\alpha(y)\linear\alpha(y+1))}_{\mbox{$w$ times}}
\linear\alpha(0)\linear\alpha(p).
$$
Functions on natural numbers can be represented by proofs
with conclusion $\judgebll{}{\Nat_x}{\Nat_p}$, where $p$ is a resource polynomial depending on $x$, only.
More generally, functions on the word algebra $\wa$ can be represented by proofs with conclusion
$\judgebll{}{\Wordalg_x}{\Wordalg_p}$. For example, all constructors $c_\wa^1,\ldots,c_\wa^w$
correspond to proofs with conclusion $\judgebll{}{\Wordalg_x}{\Wordalg_{x+1}}$, while
$\zero_\wa$ corresponds to a proof of $\judgebll{}{}{\Wordalg_0}$. More generally, the
polynomial $p$ gives a bound on the size of the result, as a function of the size of the
input. \QBAL\ supports iteration on any word algebra (including natural numbers). As an example, for every
$p$ and for every $A$ where $x$ only appears positively, there is a proof 
$\pi_p^A$ of 
$$
\judgebll{}{\Nat_p,!_{y<p}(A\{y/x\}\linear A\{y+1/x\}),A\{0/x\}}{A\{p/x\}}.
$$
namely:
{\scriptsize
$$
\infer[L_{\forall\alpha}]
{
\judgebll{}{\Nat_p,!_{y<p}(A\{y/x\}\linear A\{y+1/x\}),A\{0/x\}}{A\{p/x\}}
}
{
  \infer=[L_{\linearß†ß}]
  {\judgebll{}{!_{y<p}(A\{y/x\}\linear A\{y+1/x\})\linear A\{0/x\}\linear A\{p/x\},!_{y<p}(A\{y/x\}\linear A\{y+1/x\}),A\{0/x\}}{A\{p/x\}}}
  {
    \infer[A]
    {\judgebll{}{!_{y<p}(A\{y/x\}\linear A\{y+1/x\})}{!_{y<p}(A\{y/x\}\linear A\{y+1/x\}}}
    {}
    &
    \infer[A]
    {\judgebll{}{A\{0/x\}}{A\{0/x\}}}
    {}
    &
    \infer[A]
    {\judgebll{}{A\{p/x\}}{A\{p/x\}}}
    {}
  }
}
$$}
This will be essential to prove Lemma~\ref{lemma:contraction}.
\subsection{Unbounded First Order Quantification is Unsound} 
One may wonder why quantification on numerical variables is restricted to be
bounded (see Definition~\ref{defn:formulas}). The reason is very simple:
in presence of unbounded quantification, 
\QBAL\ would immediately become unsound.
To see that, define $\UNat$ to be the formula $\exists (x):\emptyset.\Nat_x$.
The composition of the successor with itself yields a 
proof with conclusion $\judgebll{}{\Nat_x}{\Nat_{x+2}}$ which, by rules
$R_{\exists x}$ and $L_{\exists x}$, becomes a proof with conclusion
$\judgebll{}{\UNat}{\UNat}$. Iterating it, we obtain a proof of
$\judgebll{}{\Nat_x}{\UNat}$ which represents the function $n\mapsto 2n$.
But by rule $L_{\exists x}$, it can be turned into a proof of
$\judgebll{}{\UNat}{\UNat}$, and iterating it again we obtain a proof representing
the exponential function. The boundedness assumption will be indeed critical
in Section~\ref{sect:qpt}, where we establish that any functions which
is representable in \QBAL\ is polynomial time computable.

It is not clear whether unbounded existential quantification would be
sufficient to embed the whole of second order intuitionistic logic into \QBAL.
\section{Set-Theoretic Semantics}\label{sect:sts}
\noindent In this Section, we give a set-theoretic semantics
for \QBAL. We assume that our ambient set-theory is constructive. This way we have a
set of sets $\Univ$ which contains the natural numbers, closed under binary products, function spaces and
$\Univ$-indexed products. An alternative to assuming a constructive
ambient set theory consists of replacing plain sets with PERs (partial equivalence
relations) or
domains or similar structures. See \cite{Hofmann04} for a more
detailed discussion on this issue. 

Formulas of \QBAL\ can be interpreted as sets as follows, where $\rho$ is an environment
mapping atoms to sets:
\begin{eqnarray*}
\Mean{\alpha(\multi{p})}_\rho&=&\rho(\alpha);\\
\Mean{A\otimes B}_\rho&=&\Mean{A}_\rho\times\Mean{B}_\rho;\\
\Mean{A\linear B}_\rho&=&\Mean{A}_\rho\Rightarrow\Mean{B}_\rho;\\
\Mean{\forall\alpha.A}_\rho&=&\prod_{C\in\Univ}\Mean{A}_{\rho[\alpha\mapsto C]};\\
\Mean{!_{x<p}A}_\rho=\Mean{\forall\multi{x}:\consetone.A}_\rho=\Mean{\exists\multi{x}:\consetone.A}_\rho&=&\Mean{A}_\rho.
\end{eqnarray*}
Please observe that the interpretation of any formula $A$ is completely independent from the 
resource polynomials appearing in $A$.

To any \QBAL\ proof $\pi$ of $\judgebll{\consetone}{A_1,\ldots,A_n}{B}$ we can associate a set-theoretic function
$\Mean{\pi}_\rho:\Mean{A_1\otimes\ldots\otimes A_n}_\rho\rightarrow\Mean{B}_\rho$ by induction on $\pi$, in
the obvious way. $\Mean{\pi}_\rho$ is \emph{equal} to the set-theoretic semantics of $\forget{\pi}$
as a proof of second order intuitionistic logic. Set-theoretic semantics of proofs is preserved
by cut-elimination: if $\pi$ reduces to $\sigma$ by cut-elimination, then $\Mean{\pi}_\rho=\Mean{\sigma}_\rho$.

Observe that $\Mean{A}_\rho$ only depends on the values of $\rho$ on atoms appearing free in $A$.
So, in particular,
$$
\Mean{\Nat_q}_\rho=\prod_{C\in\Univ}(C\Rightarrow C)\Rightarrow(C\Rightarrow C)
$$
is independent on $\rho$ and on $q$, since $\Nat_q$ is a closed formula. Similarly for $\Mean{\Wordalg_q}_\rho$.
Actually, there are functions $\varphi_\N:\N\rightarrow\Mean{\Nat_p}$ and
$\psi_\N:\Mean{\Nat_p}\rightarrow\N$ such that $\psi_\N\circ\varphi_\N$ is the
identity on natural numbers. They are defined as follows:
\begin{eqnarray*}
(\varphi_\N(n))_C(f,z)&=&f^n(z);\\
\psi_\N(x)&=&x_\N(x\mapsto x+1)(0);
\end{eqnarray*}
where $A_C$ is the projection of $A$ on the component $C$ whenever $A$ is some product
$\prod_{D\in\Univ}B$.

So, given a proof $\pi:\judgebll{}{\Nat_x}{\Nat_p}$, the \emph{numeric function represented by} $\pi$
is simply $\psi_\N\circ\Mean{\pi}\circ\varphi_\N$. Similar arguments hold for
functions with conclusion $\judgebll{}{\Wordalg_x}{\Wordalg_p}$.
\section{\QBAL\ and Polynomial Time}\label{sect:qpt}
\noindent In this section we show that all functions on natural
numbers definable in \QBAL\ are polynomial time computable. To this end, we follow the
semantic approach in \cite{Hofmann04} which we now summarizes. 
\subsection{Realizability Sets}
Let $X$ be a finite set of resource variables. We write $\V(X)$ for
$\N^X$ --- the elements of $\V(X)$ are called \emph{valuations} (over
$X$). If $\eta\in\V(X)$ and $c\in\N$ then $\eta[x\mapsto c]$ denotes
the valuation which maps $x$ to $c$ and acts like $\eta$ otherwise.
We assume some reasonable encoding of valuations as natural numbers allowing
them to be passed as arguments to algorithms.

If $\consetone$ is a constraint set involving at most the variables in
$X$ (\emph{over} $X$) then $\V_\consetone(X)$ (or simply $\V_{\consetone}$)
is the set of valuations in $\V(X)$ satisfying all the constraints
in $\consetone$.

We write $\P(X)$ for the set of resource polynomials over $X$. If
$p\in\P(X)$ and $\eta\in\V(X)$ we write $p(\eta)$ for the number
obtained by evaluating $p$ with $x\mapsto \eta(x)$ for each $x\in X$.
A \emph{substitution} $\sigma:X\rightarrow Y$ is a function
mapping any variable in $Y$ to a polynomial in $\P(X)$. Given a substitution
$\sigma:X\rightarrow Y$ and a valuation $\eta\in\V(X)$, the valuation
$\sigma[\eta]\in\V(Y)$ assigns to every variable $y\in Y$ the 
natural number $\sigma(y)(\eta)$.

We assume known the untyped lambda calculus as defined e.g.\ in
\cite{Barendregt84}. 
A lambda term is \emph{affine linear} if each variable (free
or bound) 
appears at most once (up to $\alpha$-congruence). For example, $\lambda x.\lambda y.yx$ and
$\lambda x.\lambda y.y$ and $\lambda x.xy$ are affine linear while the term $\lambda x.xx$ is
not. Notice that every affine linear term $t$ is strongly normalisable in less than
$|t|$ steps where $|t|$ is the size of the term. Moreover, the size $|s|$ of any
reduct of $t$ is at most $|t|$. The runtime of the computation leading to the normal form is therefore 
$O(|t|^2)$.  We will henceforth use the expression \emph{affine linear term} for an affine linear
lambda term which is in normal form. If $s,t$ are affine linear terms,
then their application $st$ is defined as the normal form of the
lambda term $st$. Notice that the application $st$ can be computed in
time $O((|s|+|t|)^2)$. 

If $s,t$ are affine linear terms we write $s{\otimes}t$ for the affine
linear term $\lambda f.fst$. If $t$ is an affine linear term possibly
containing the free variables $x,y$ then we write $\lambda
x{\otimes}y.t$ for $\lambda u.u(\lambda x\lambda y.t)$. Notice that
$(\lambda x{\otimes}y.t)(u{\otimes}v)=t\{u/x,v/y\}$.  

More generally, if $(t_i)_{i<n}$ is a family of affine linear terms,
we write $\bigotimes_{i<n}t_i$ and
$\lambda \bigotimes_{i<n}x_i.t$ for $\lambda f.ft_0 t_1\dots t_{n-1}$,
respectively, $\lambda u.u(\lambda x_0\lambda
x_1\dots\lambda x_{n-1}.t)$. Again, 
$$
(\lambda\bigotimes_{i<n}x_i.t)
(\bigotimes_{i<n}t_i)=
t\{t_0/x_0,\ldots,t_{n-1}/x_{n-1}\}.
$$
We write $\LA$ for the set of closed affine linear terms. 

There is a canonical way of representing terms of any word
algebra $\wa$ as affine linear terms, which is attributed to
Dana Scott~\cite{Wadsworth80}. If the unary constructors of the word algebra $\wa$
are $c_\wa^1,\ldots,c_\wa^w$ and $\zero_\wa$ is the only $0$-ary constructor of $\wa$, 
the terms of $\wa$ are mapped to affine linear terms as follows:
\begin{eqnarray*}
  \encode{\zero_\wa}&=\lambda x_1.\ldots\lambda x_w.\lambda y.y;\\
  \forall i\in\{1,\ldots,w\}.\encode{c_\wa^is}&=\lambda x_1.\ldots\lambda x_w.\lambda y.x_i\encode{s}.
\end{eqnarray*}
As an example, the natural number $2$ seen as a term of the word algebra
$\bN$ becomes
$$
\encode{2}=\lambda x.\lambda y.x(\lambda x.\lambda y.x(\lambda x.\lambda y.y))
$$
\begin{defi}[Realizability Set]
  Let $X$ be a finite set of resource variables.  A
  \emph{realizability set over $X$} is a pair $A = (|A|,\forces{A})$
  where $|A|$ is a set and $\forces{A}\subseteq\V(X)\times\LA\times
  |A|$ is a ternary relation between valuations over $X$, affine
  lambda terms, and the set $|A|$.  We write $\eta,t\forces{A} a$ for
  $(\eta,t,a)\in\forces{A}$. Given a substitution $\sigma$ from $X$ to
  $Y$ and a realizability set $A$ over $Y$, then a new 
  realizability set $A[\sigma]$ over $X$ is defined by 
  $|A[\sigma]|=|A|$ and $\eta,t\forces{A[\sigma]} a$
  iff $\sigma[\eta],t\forces{A} a$.
\end{defi}
The intuition behind $\eta,t\forces{A} a$ is that $a$ is an
abstract semantic value, $\eta$ measures the abstract
size of $a$, and the affine linear term $t$ encodes the abstract value
$a$. This is a generalization of what normally happens in realizability
models, where $\forces{A}$ is a \emph{binary} relation between realizers
and denotations.
\begin{exa}\label{nexpl}\hfill
\begin{enumerate}[$\bullet$]
\item
  The realizability set $\bN_x$ over $\{x\}$  of {\em tally natural
  numbers}  (``of size at most $x$'') is 
  defined  by:  $|\bN_x|=\N$  and  
  $\eta,t\forces{\bN_x} n$ if $t=\encode{n}$ and
  $\eta(x)\geq n$;
\item
  The realizability set $\Wordalg_x$ over $\{x\}$ of {\em free terms of $\wa$}
  (``of length at most $x$'') is 
  defined  by:  $|\Wordalg_x| = \wa$ and 
  $\eta,t\forces{\Wordalg_x} w$ if $t=\ulcorner w \urcorner$
  and $\eta(x)\geq |w|$.  
\end{enumerate}
\end{exa}

\noindent These realizability sets $\bN_x$ and $\Wordalg_x$ 
turn out to be retracts of the denotations of the
eponymous \BLL\ formulas from Section~\ref{sect:programming}.
\begin{defi}[Positive and Negative Variables]
Let $A$ be a realizability set over $X$. We say that $x\in X$ is
\emph{positive} (\emph{negative}, respectively) in $A$, if for all $\eta,\mu\in\V(X),
t\in\LA,a\in|A|$ where $\eta$ and $\mu$ agree on $X\setminus\{x\}$
and $\eta(x)\leq\mu(x)$ ($\eta(x)\geq\mu(x)$, respectively),
$\eta,t\forces{A}a$ implies
$\mu,t\forces{A}a$. 
\end{defi}
We notice that $x$ is positive in $\bN_x$ and $\Wordalg_x$. Indeed, if e.g.
$\eta,t\forces{\bN_x} n$ and $\eta(x)\leq\mu(x)$, then
$\mu(x)\geq \eta(x)\geq n$ and 
$\mu,t\forces{\bN_x} n$ by definition.

Realizability sets can be thought of as the object of a category
whose arrows are functions, themselves realized by affine linear
terms, one for every possible valuation of the underlying resource
variables:
\begin{defi}[Morphisms]
Let $A,B$ be realizability sets over some set $X$. 
A {\em morphism} from $A$ to $B$ is a function
$f:|A|\ra |B|$ satisfying the following conditions:
\begin{enumerate}[$\bullet$]
\item 
  there exist a function $e:\V(X)\rightarrow \LA$, an algorithm $A$ and a resource
  polynomial $q$ such that for every $\eta\in\V(X)$, $A$ computes
  $e(\eta)$ in time bounded by $q(\eta)$;
\item
  for each $\eta\in\V(X)$, $t\in\LA$, $a\in|A|$, we have
  $$
  \eta,t\forces{A}a\quad\mbox{implies}\quad\eta,e(\eta)t\forces{B}f(a).
  $$
\end{enumerate}
In this case we say that $e$ \emph{witnesses} $f$ and write 
$A\hmap{f}{e}B$ where in the notation the algorithm $A$ computing $e$ is presumed to exist.
\end{defi}
Noticeably, morphisms compose.

The following definitions summarises the interpretation of formulas according to \cite{Hofmann04}.
First of all, multiplicative connectives $\otimes$ and $\linear$ correspond to
constructions on realizability sets:
\begin{defi}[Multiplicatives] 
Let $A, B$ be realizability sets over $X$. Then the following are realizability
sets over $X$:
\begin{enumerate}[$\bullet$]
\item 
  $A\otimes B$ as given 
  by $|A\otimes B|=|A|\times|B|$ and 
  $\eta,t\forces{A\otimes B}(a,b)$ iff $t=u{\otimes}v$,
  where $\eta,u \forces{A} a$ and $\eta,v \forces{B} b$.
\item  
  $A\limp B$ is given by $|A\limp B| = |A|\Rightarrow |B|$
  and $\eta,t\forces{A\limp B} f$ iff whenever $\eta,u\forces{A} a$
  it holds that $\eta,t\,u\forces{B} f(a)$. 
\end{enumerate}
\end{defi}
Another logical connective needs to be justified, namely the exponential modality:
\begin{defi}[Modality]
If $C$ is a realizability set over $X\cup\{x\}$ and $p\in\P(X)$ then a realizability set $!_{x<p}C$ over $X$ is defined by
$|\bang{}_{x<p}C|=|C|$ and  $\eta,t\forces{\bang{}_{x<p} C} a$ if 
\begin{enumerate}[$\bullet$]
\item $t=\bigotimes_{i<p(\eta)} t_i$ for some family
  $(t_i)_{i<p(\eta)}$;
\item $\eta[x\mapsto i],t_i\forces{C} a$ for each $i<p(\eta)$. 
\end{enumerate}
\end{defi}
Lastly, a semantical counterpart of second order universal quantification
must be defined. The following are essential preliminary definitions.
\begin{defi}[Second Order Environments]
Let $X$ be a set of resource variables. A \emph{second-order environment over 
$X$} is a partial function $\rho$ 
which assigns to a second-order variable $\alpha$ of arity $n$ a pair 
$(l,C)$ such that:
\begin{enumerate}[$\bullet$]
  \item
    $l=(y_1,\ldots,y_n)$ is an $n$-tuple of pairwise different
    resource variables not occurring in $X$;
  \item
    $C$ is a realizability set over $X\cup\{y_1,\ldots, y_n\}$ 
    in which the $y_i$ are positive.
\end{enumerate} 
For a second-order environment $\rho$ we write 
$|\rho|$ for the mapping $\alpha\mapsto |C|$ when 
$\rho(\alpha)=(l,C)$.
If $\sigma:X\rightarrow Y$ is a substitution and 
$\rho$ is a second-order environment over $Y$ we define a 
second-order environment $\rho[\sigma]$ over $X$ by 
$\rho[\sigma](\alpha)= (l,C[\sigma])$ when $\rho(\alpha)=(l,C)$. 
We assume here that the variables in $l$ are not contained in $Y$. 
Otherwise, the substitution cannot be defined.
\end{defi}

Using these semantic constructions one defines for each formula $A$
with free resource variables contained in $X$ and second-order
environment $\rho$ over $X$, a realizability set $\MeanB{A}_\rho$ over
$X$ in such a way that $|\MeanB{A}_\rho| = \Mean{A}_{|\rho|}$ (where
$|\rho|$ is the assignment of sets to atoms obtained from $\rho$ in the
obvious way), that is to say, the underlying set of the realizability set interpreting a
formula $A$ coincides with the set-theoretic meaning of $A$ (see Section~\ref{sect:sts}):
$$
\MeanB{\alpha(p_1,\ldots,p_n)}_\rho=C[\sigma],
$$
where $\rho(\alpha)=((y_1,\ldots,y_n),C)$ and $\sigma(y_i)=p_i$;
\begin{eqnarray*}
\MeanB{A\otimes B}_\rho&=\MeanB{A}_\rho\otimes\MeanB{B}_\rho;\\
\MeanB{A\linear B}_\rho&=\MeanB{A}_\rho\linear\MeanB{B}_\rho;\\
\MeanB{\forall\alpha.A}_\rho&=\left(\prod_{C\in\Univ}\Mean{A}_{|\rho|[\alpha\mapsto C]},\forces{\MeanB{\forall\alpha.A}_\rho}\right);
\end{eqnarray*}
where $\eta,t\forces{\MeanB{\forall\alpha.A}_\rho}f$ iff
$\eta,t\forces{\MeanB{A}_{\rho[\alpha\mapsto(l,C)]}}f_C$ for all $(l,C)$;
$$
\MeanB{!_{x<p}A}_\rho=!_{x<p}\MeanB{A}_{\rho[\weak{x}]},
$$
where $\weak{x}:X\cup\{x\}\rightarrow X$ is the substitution mapping
any variable in $X$ into the same variable as an element of
$\P(X\cup\{x\})$.

The main result of \cite{Hofmann04} then asserts that if $\pi$ is a
proof (in \BLL) of a sequent $\Gamma\vdash B$ then the function
$\Mean{\pi}_{|\rho|}$ is a morphism from $\MeanB{\Gamma}_\rho$ to
$\MeanB{B}_\rho$ (where we interpret a context $\Gamma$ as a
$\otimes$-product over its components as usual).  From this, polynomial
time soundness is a direct corollary since polynomial time
computability is built into the notion of a morphism.

It thus only remains to extend the realizability model to cover the
constructs of \QBAL{} which we do in the next section. 
\subsection{Extending the Realizability Model to \QBAL}
The notion of a realizability set above is adequate to model formulas of \QBAL. The
notion of a morphism, however, should be slightly generalized in order to capture
constraints:
\begin{defi}[$\consetone$-Morphisms]
Let $A,B$ be realizability sets over some set $X$ and $\consetone$ a constraint set over $X$.
A function $f:|A|\rightarrow |B|$ is a $\consetone$-morphism from $A$ to $B$ iff the following
conditions hold:
\begin{enumerate}[$\bullet$]
\item 
  there exist a function $e:\V_\consetone(X)\rightarrow \LA$ and an algorithm $A$ such that
  $A$ computes $e(\eta)$ from $\eta$ in time bounded by $q(\eta)$ for some resource polynomial
  $q$;
\item
  for each $\eta\in\V_\consetone(X)$, $t\in\LA$, $a\in|A|$, we have
  that $\eta,t\forces{A}a$ implies $\eta,e(\eta)t\forces{B}f(a)$.
\end{enumerate}
\end{defi}

\noindent In order to define realizability sets $\forall\multi{y}{:}\consetone.A$
and $\exists\multi{y}{:}\consetone.A$, we fix some encoding of environments $\eta$ as affine
lambda terms using the $\ulcorner\cdot\urcorner$ encoding of natural
numbers. As an example, the environment $\eta$ on $\{x_0,\ldots,x_{n-1}\}$ could be encoded
as $\bigotimes_{i<n}\encode{\eta(x_i)}$; this clearly relies on a total order on resource
variables. We do not notationally distinguish environments from their encodings.
\begin{defi}[First-order Quantification]\label{kli}
Let $X,Y$ be disjoint sets of variables. 
Let $A$ be a realizability set over $X\cup Y$ and $\consetone$ a constraint set over $X\cup Y$ 
where we put $Y=\{y_1,\dots,y_n\}$ and $\multi{y}=(y_1,\dots,y_n)$. Furthermore, for each $i=1,\dots,n$ let $p_i\in \P(X)$ 
be such that $\consetone\models\{\multi{y}\leq \multi{p}\}$.
\begin{enumerate}[$\bullet$]
\item $|\forall \multi{y}{:}\consetone.A|=|\exists \multi{y}{:}\consetone.A|=|A|$, 
\item $\eta,t\forces{\forall \multi{y}{:}\consetone.A} a\iff 
 \forall {\mu}\in\V(Y). \eta{\cup}{\mu}\in\V_\consetone\Rightarrow 
       \eta{\cup}{\mu},t{\mu}\forces{A}a$. 
\item $\eta,\mu\otimes t\forces{\exists \multi{y}{:}\consetone.A} a\iff 
       \mu\in\V(Y)\wedge \eta{\cup}{\mu}\in\V_\consetone\wedge
         \eta{\cup}{\mu},t\forces{A}a$
\end{enumerate}
\end{defi}       
Recall that $\forall\multi{y}{:}\consetone.A$ and
$\exists\multi{y}{:}\consetone.A$ are well-formed only if
there are resource polynomials $\multi{p}$ such that
$\consetone\models \multi{y}<\multi{p}$. Therefore, 
 the set $\{{\mu}\mid
\eta{\cup}{\mu}\in\V_\consetone\}$ is finite and in fact computable
in polynomial time from $\eta$. Indeed, its cardinality
at most
$$
\prod_{i=1}^np_i(\eta),
$$
and the size of any of its elements is at most
$$
|\eta|+\prod_{i=1}^np_i(\eta).
$$

We are now able to prove the main result of this Section:
\begin{thm}
Let $\pi$ be a proof of a sequent
$\judgebll{\consetone}{\Gamma}{B}$ and $\rho$ a mapping of atoms to
  realizability sets. Then $\Mean{\pi}_{|\rho|}$ is a $\consetone$-morphism from
  $\MeanB{\Gamma}_\rho$ to $\MeanB{B}_\rho$. 
\end{thm}
\proof
The proof is by induction on derivations. 
We only show the cases that differ significantly from the development in \cite{Hofmann04}.

\noindent
\emph{Case $P_!$.} For simplicity, suppose that $n=1$, $q_1=p$ and $A_1=A$. 
The induction hypothesis shows
that $\Mean{\pi}_{|\rho|}$ is a $\consetone$-morphism from
$\MeanB{A}_\rho$ to $\MeanB{B}_\rho$ witnessed by $e$. As in the
proof of the main result in \cite{Hofmann04}, we define
$$
d(\eta)=\lambda\bigotimes_{i< p(\eta)}x_i.\bigotimes_{i< p(\eta)}e(\eta[x\mapsto i])x_i.
$$
Now, if $\eta\in\V_\consettwo$,
then $\eta[x\mapsto i]\in\V_\consetone$ whenever $i< p(\eta)$ by the side condition from rule $P_!$.
We obtain that $\Mean{\pi}$ is a $\consettwo$-morphism from 
$\MeanB{!_{x{<}p}A}_\rho$ to $\MeanB{!_{x{<}p}B}_\rho$ witnessed by $d$. 

\noindent
The remaining cases are the four rules for first order quantifiers. In
each case, we assume by the induction hypothesis that $\Mean{\pi}$ is
a morphism realizing the premise of the rule and let $e$ be its witness. We
have to show that $\Mean{\pi}$ is a morphism realizing the conclusion
of the rule. Note that the set-theoretic meaning of a proof does not
change upon application of any of the quantifier rules. 

\noindent
\emph{Case $R_{\forall x}$.}
Suppose that $\eta\in\V_\consetone$ and
$\eta,t_\gamma\forces{\MeanB{\Gamma}_\rho} \gamma$. Now suppose
$\eta\cup \mu\in\V_\consettwo$. By the induction hypothesis
 $\eta{\cup}\mu,e(\eta{\cup}\mu)t_\gamma\forces{\MeanB{A}_\rho}
 \Mean{\pi}(\gamma)$. We thus define $d$ by 
$d(\eta)=u$
where $u\in\LA$ is such that $ut\mu=e(\eta\cup\mu)t$ whenever
$\eta\cup\mu\in\V_\consetone$. Recall that for a given $\eta$
there are only $q(\eta)$ such $\mu$ (for a fixed resource polynomial
$q$), so that $t$ can be constructed
as a big case distinction over all those $\mu$.
It is then clear that $d$ is polynomial time computable and realizes the
conclusion of the rule. 

\noindent
\emph{Case $L_{\forall x}$.}
Assume $\eta\in\V_\consetone$ and $\eta,t_\gamma\forces{\MeanB{\Gamma}_\rho} \gamma$ and 
$\eta,t_a\forces{\MeanB{\forall \multi{x}:\consettwo.A}_\rho} a$. 
Define $\mu_\eta\in\V(X)$ by
$\mu_\eta(x_i)=p_i(\eta)$ so that $\eta\cup\mu_\eta\in\V_\consettwo$ by the side condition to the rule. 
Now, $\eta{\cup}\mu_\eta,t_a{\mu_\eta}\forces{A} a$ by
Definition~\ref{kli}. Hence,
$\eta,t_a{\mu_\eta}\forces{A\{\multi{p}/\multi{x}\}}a$. By the induction
hypothesis, $e(\eta)(t_\gamma\otimes t_a{\mu_\eta})\forces{C} \Mean{\pi}(\gamma,a)$, so
$d(\eta)=\lambda x_\gamma\otimes x_a.e(\eta)(x_\gamma\otimes 
x_a (\mu_\eta))$ does the job. 
\noindent
The remaining two cases are essentially dual to the previous two. 
We merely define the witnesses. 

\noindent
\emph{Case $R_{\exists x}$.}
Define $\mu(\eta)$ as in $L_{\forall x}$. We can then put 
$d(\eta)=\lambda t_\gamma.\mu(\eta)\otimes e(\eta)t_\gamma$. 

\noindent
\emph{Case $L_{\exists x}$.}
We define $d(\eta)$ to be such that 
$d(\eta)t_\gamma (\mu\otimes t)=e(\eta\cup\mu)t_\gamma t$. This is possible by hard-wiring 
separate cases for each of the polynomially in $\eta$ many $\mu$ like in case $R_{\forall x}$.\qed

\begin{cor}
Every function on word algebras representable in $\QBAL$ is polynomial time computable.
\end{cor}

\section{On Compositional Embeddings}\label{sect:oce}
\noindent In this Section, we justify our emphasis on compositional embeddings.
An embedding of a logical system or programming language $\Ls$ into $\QBAL$
is a function $\trlqbllpr{\cdot}$ from the space of proofs (or programs) of $\Ls$ into the space
of proofs for $\QBAL$. Clearly, for an embedding to be relevant from a computational
point of view, any proof $\pi$ of $\Ls$ should be mapped to an equivalent 
proof $\trlqbllpr{\pi}$, e.g., $\Mean{\trlqbllpr{\pi}}=\Mean{\pi}$. The existence
of an embedding of $\Ls$ into $\QBAL$ implicitly proves that $\QBAL$ is \emph{extensionally}
at least as powerful as $\Ls$. Such an embedding $\trlqbllpr{\cdot}$ is not necessarily
computable nor natural. But whenever $\Ls$ is a sound and complete ICC characterization
of polynomial time, a large class of proofs or programs of $\Ls$ can be mapped to \QBAL, since the 
classes of definable \emph{first order} functions are exactly the same in $\Ls$ and \QBAL. 
Indeed, \QBAL\ is both extensionally sound (see Section~\ref{sect:qpt})
and extensionally complete (since \BLL\ can be compositionally embedded into it).

Typically, one would like to go beyond extensionality and
prove that $\QBAL$ is \emph{intensionally}
as powerful as $\Ls$. And if this is the goal, $\trlqbllpr{\cdot}$ should be
easily computable. Ideally, we would like $\trlqbllpr{\cdot}$ 
to act homeomorphically on the space of proofs of $\Ls$. In other words, whenever
a proof $\pi$ of $\Ls$ is obtained applying a proof-forming rule $R$ to $\rho_1,\ldots,\rho_n$,
then $\trlqbllpr{\pi}$ should be obtainable from $\trlqbllpr{\rho_1},\ldots,\trlqbllpr{\rho_n}$
in a uniform way, i.e., dependently on $R$ but independently on 
$\trlqbllpr{\rho_1},\ldots,\trlqbllpr{\rho_n}$. An embedding satisfying the above constraint
is said to be \emph{strongly compositional}. The embeddings we will present
in the following two sections are only \emph{weakly} compositional: 
$\forget{\trlqbllpr{\pi}}$ can be uniformly built from $\forget{\trlqbllpr{\rho_1}},\ldots,\forget{\trlqbllpr{\rho_n}}$
whenever $\pi$ is obtained applying $R$ to $\rho_1,\ldots,\rho_n$.
We believe that the existence of a weakly compositional embedding of $\Ls$ into $\QBAL$ is
sufficient to guarantee that $\QBAL$ is intensionally as powerful as $\Ls$ because,
as we pointed out in Section~\ref{sect:qbllsol}, $\forget{\pi}$ can be thought as the program hidden
in the proof $\pi$.
\section{Embedding \LFPL}\label{sect:emlfpl}
\noindent \LFPL\ is a calculus for non-size-increasing computation introduced by the
second author~\cite{Hofmann03}. It allows to capture natural algorithms computing
functions such that the size of the result is smaller or equal to the size of the
arguments. This way, polynomial time soundness is guaranteed despite the possibility
of arbitrarily nested recursive definitions.

We here show that a core subset of \LFPL\ can be compositionally embedded into
\QBAL. \LFPL\ types are generated by the following grammar:
$$
A::=\diamond\spb\nat\spb A\otimes A\spb A\linear A.
$$
Rules for \LFPL\ in natural-deduction style are in Figure~\ref{figure:lfplrules}.
We omit terms, since the computational content of type derivations is implicit
in their skeleton.
\begin{figure*}[htbp]
\begin{center}
\fbox{
\begin{minipage}{.97\textwidth}
\centering\textbf{Axiom, Base Types and Weakening}
$$
\begin{array}{ccccccc}
\infer[A]
{A\vdash A}
{}
&\hspace{4mm}&
\infer[S]
{\diamond,\nat\vdash\nat}
{}
&\hspace{4mm}&
\infer[T]
{\vdash\nat\linear A}
{\diamond\vdash A\linear A & \vdash A}
&\hspace{4mm}&
\infer[W]
{\Gamma,B\vdash A}
{\Gamma\vdash A}
\end{array}
$$
\centering\textbf{Multiplicative Rules}
$$
\begin{array}{ccc}
\infer[I_\linear]
{\Gamma\vdash A\linear B}
{\Gamma,A\vdash B} 
&\hspace{8mm}&
\infer[E_\linear]
{\Gamma,\Delta\vdash B} 
{\Gamma\vdash A\linear B & \Delta\vdash A}
\end{array}
$$
$$
\begin{array}{ccc}
\infer[I_\otimes]
{\Gamma,\Delta\vdash A\otimes B}
{\Gamma\vdash A & \Delta\vdash B} 
&\hspace{4mm}&
\infer[E_\otimes]
{\Gamma,\Delta\vdash C} 
{\Gamma\vdash A\otimes B & \Delta,A,B\vdash C}
\end{array}
$$
\end{minipage}}
\caption{\LFPL}
\label{figure:lfplrules}
\end{center}
\end{figure*}
The set-theoretic semantics $\Mean{A}$ of an \LFPL\ formula $A$ can be defined very easily:
$\Mean{\diamond}=\prod_{C\in\Univ}C\Rightarrow C$,
$\Mean{\Nat}=\prod_{C\in\Univ}(C\Rightarrow C)\Rightarrow (C\Rightarrow C)$, while
the operators $\otimes$ and $\linear$ are interpreted as usual. Notice
that the interpretation of an \LFPL\ formula does not depend on any environment
$\rho$. This way, any
\LFPL\ proof $\pi:A_1,\ldots,A_n\vdash B$ can be given a semantics
$\Mean{\pi}:\Mean{A_1\otimes\ldots\otimes A_n}\rightarrow\Mean{B}$, itself independent
on any $\rho$. For example, rule $T$ corresponds to iteration, while rule $E_\linear$ 
corresponds to function application.

\LFPL\ types can be translated to \QBAL\ formulas in the following way:
\begin{eqnarray*}
\trlfplqbll{\diamond}{p}{q}&=&\exists\varepsilon:\{1\leq p\}.\forall\alpha.\alpha\linear\alpha\\
\trlfplqbll{\nat}{p}{q}&=&\natbll{p}\\
\trlfplqbll{A\otimes B}{p}{q}&=&\exists(x,y):\{x+y \leq p\}.\trlfplqbll{A}{x}{q}\otimes\trlfplqbll{B}{y}{q}\\
\trlfplqbll{A\linear B}{p}{q}&=&\forall(x):\{x+p\leq q\}.\trlfplqbll{A}{x}{q}\linear\trlfplqbll{B}{p+x}{q}
\end{eqnarray*}
Please observe that the interpretation of any \LFPL\ formulas is parametrized by two resource polynomials $p$ and
$q$. If a variable $x$ occurs in $p$, but not in $q$, then $x$ occurs only positively in 
$\trlfplqbll{A}{p}{q}$: this can be proved by an easy induction on the structure of $A$.

The correspondence scales from types to proofs:
\begin{thm}\label{thm:lfpl}
\LFPL\ can be embedded into \QBAL. In other words,
for every \LFPL\ proof
$\pi:A_1,\ldots,A_n\vdash B$, there exists a \QBAL\ proof 
$$
\trlfplqbllpr{\pi}:
\judgebll{\{\sump{i}{x}\leq b,1\leq b\}}{\trlfplqbll{A_1}{x_1}{b},\ldots,\trlfplqbll{A_n}{x_n}{b}}{\trlfplqbll{B}{\sump{i}{x}}{b}}
$$
such that $\Mean{\pi}=\Mean{\trlfplqbllpr{\pi}}$. 
\end{thm}
\proof
As expected, the proof goes by induction on $\pi$.
\begin{enumerate}[$\bullet$]
\item
  If the only rule in $\pi$ is $A$, then
  $\trlfplqbllpr{\pi}$ is simply the axiom
  $$
  \infer[A]
  { \judgebll{\{x\leq b,1\leq b\}}{\trlfplqbll{A}{x}{b}}{\trlfplqbll{A}{x}{b}}}
  {}
  $$
\item
  If the only rule in $\pi$ is $S$, then
  $\trlfplqbllpr{\pi}$ is
  $$
  \infer[L_{\exists x}]
  {
    \judgebll{\{x+y\leq b,1\leq b\}}{\trlfplqbll{\diamond}{x}{b},\trlfplqbll{\nat}{y}{b}}
    {\trlfplqbll{\nat}{x+y}{b}}
  }
  {
    \infer[W]
    {
      \judgebll{\{x+y\leq b,1\leq b,1\leq x\}}{\forall\alpha.\alpha\linear\alpha,\trlfplqbll{\nat}{y}{b}}{\trlfplqbll{\nat}{x+y}{b}}
    }
    {
      \rho:\judgebll{\{x+y\leq b,1\leq b,1\leq x\}}{\trlfplqbll{\nat}{y}{b}}{\trlfplqbll{\nat}{x+y}{b}}
    }
  }
  $$
  where $\sigma:\judgebll{}{\trlfplqbll{\nat}{y}{b}}{\trlfplqbll{\nat}{y+1}{b}}$ is the \QBAL\ proof for the
  successor on natural numbers inherited from \BLL\ and $\rho$ is obtained by first strengthening
  $\sigma$ into a proof of $\judgebll{\{x+y\leq b,1\leq b,1\leq x\}}{\trlfplqbll{\nat}{y}{b}}{\trlfplqbll{\nat}{y+1}{b}}$
  (by Proposition~\ref{prop:strengthcon}) and then applying to it Proposition~\ref{prop:monotonicity}, after observing
  that $\trlfplqbll{\nat}{y+1}{b}\leform_{\{x+y\leq b,1\leq b,1\leq x\}}\trlfplqbll{\nat}{y+x}{b}$.
\item
  If the last rule in $\pi$ is $W$ and the immediate premise of $\pi$ is $\rho$, then 
  $\trlfplqbllpr{\pi}$ is
  $$
  \infer[W]
  {\judgebll{\{\sump{i}{x}+y\leq b,1\leq b\}}
      {\trlfplqbll{\Gamma}{\multi{x}}{b},\trlfplqbll{A}{y}{b}}
      {\trlfplqbll{B}{\sump{i}{x}+y}{b}}}
  {\sigma:\judgebll{\{\sump{i}{x}+y\leq b,1\leq b\}}
      {\trlfplqbll{\Gamma}{\multi{x}}{b}}
      {\trlfplqbll{B}{\sump{i}{x}+y}{b}}}
  $$
  where $\sigma$ can be obtained from $\trlfplqbllpr{\rho}$ by Proposition~\ref{prop:strengthcon}
  and Proposition~\ref{prop:monotonicity}, because
  $\{\sump{i}{x}+y\leq b,1\leq b\}\models \{\sump{i}{x}\leq b,1\leq b\}$ and
  $\trlfplqbll{B}{\sump{i}{x}}{b}\leform \trlfplqbll{B}{\sump{i}{x}+y}{b}$
\item
  If the last rule in $\pi$ is $E_\linear$ and the immediate premises of $\pi$ are $\rho$ and $\sigma$, then
  $\trlfplqbllpr{\pi}$ is  
  {\scriptsize
  $$
  \infer=[\!L_\linear,A,U]
  {
    \judgebll{\{\sump{i}{x}+\sump{i}{y}\leq b,1\leq b\}}{\trlfplqbll{\Gamma}{\multi{x}}{b},\trlfplqbll{\Delta}{\multi{y}}{b}}
    {\trlfplqbll{B}{\sump{i}{x}+\sump{i}{y}}{b}}
  }
  {
    \infer=[\!L_{\forall x},A,U]
    {
      \judgebll{\{\sump{i}{x}+\sump{i}{y}\leq b,1\leq b\}}{\trlfplqbll{\Gamma}{\multi{x}}{b}}
      {\trlfplqbll{A}{\sump{i}{y}}{b}\linear\trlfplqbll{B}{\sump{i}{x}+\sump{i}{y}}{b}}
    }
    {
      \theta:\judgebll{\{\sump{i}{x}+\sump{i}{y}\leq b,1\leq b\}}{\trlfplqbll{\Gamma}{\multi{x}}{b}}
      {\trlfplqbll{A\linear B}{\sump{i}{x}}{b}}
    }
    &
    \!\!\!\xi:\judgebll{\{\sump{i}{x}+\sump{i}{y}\leq b,1\leq b\}}{\trlfplqbll{\Delta}{\multi{y}}{b}}
    {\trlfplqbll{A}{\sump{i}{y}}{b}}
  }
  $$}
  where $\theta$ and $\xi$ can be obtained from $\trlfplqbllpr{\rho}$ and $\trlfplqbllpr{\sigma}$,
  respectively, by applying Proposition~\ref{prop:strengthcon}.
\item
  If the last rule in $\pi$ is $I_\linear$ and the immediate premise of $\pi$ is $\rho$, then 
  $\trlfplqbllpr{\pi}$ is  
  $$
  \infer[R_{\forall x}]
  {
    \judgebll{\{\sump{i}{x}\leq b,1\leq b\}}{\trlfplqbll{\Gamma}{\multi{x}}{b}}{\trlfplqbll{A\linear B}{\sump{i}{x}}{b}}
  }
  {
    \infer[R_\linear]
    {
      \judgebll{\{\sump{i}{x}\leq b,\sump{i}{x}+y\leq b,1\leq b\}}
      {\trlfplqbll{\Gamma}{\multi{x}}{b}}
      {\trlfplqbll{A}{y}{b}\linear\trlfplqbll{B}{\sump{i}{x}+y}{b}}
    }
    {
      \sigma:\judgebll{\{\sump{i}{x}\leq b,\sump{i}{x}+y\leq b,1\leq b\}}
      {\trlfplqbll{\Gamma}{\multi{x}}{b},\trlfplqbll{A}{y}{b}}
      {\trlfplqbll{B}{\sump{i}{x}+y}{b}}
    }
  }
  $$
  where $\sigma$ can be obtained from $\trlfplqbllpr{\rho}$ by applying Proposition~\ref{prop:strengthcon}.
\item
  If the last rule in $\pi$ is $I_\otimes$ and the immediate premises of $\pi$ are $\rho$ and $\sigma$, then
  $\trlfplqbllpr{\pi}$ is  
  $$
  \infer[R_{\exists x}]
  {
    \judgebll{\{\sump{i}{x}\leq b,\sump{i}{y}\leq b,1\leq b\}}{\trlfplqbll{\Gamma}{\multi{x}}{b},\trlfplqbll{\Delta}{\multi{y}}{b}}
    {\trlfplqbll{A\otimes B}{\sump{i}{x}+\sump{i}{y}}{b}}
  }
  {
    \infer[R_\otimes]
    {
      \judgebll{\{\sump{i}{x}\leq b,\sump{i}{y}\leq b,1\leq b\}}{\trlfplqbll{\Gamma}{\multi{x}}{b},\trlfplqbll{\Delta}{\multi{y}}{b}}
      {\trlfplqbll{A}{\sump{i}{x}}{b}\otimes\trlfplqbll{B}{\sump{i}{y}}{b}}
    }
    {
      \theta:\judgebll{\{\sump{i}{x}\leq b,\sump{i}{y}\leq b,1\leq b\}}{\trlfplqbll{\Gamma}{\multi{x}}{b}}
      {\trlfplqbll{A}{\sump{i}{x}}{b}}
      &
      \xi:\judgebll{\{\sump{i}{x}\leq b,\sump{i}{y}\leq b,1\leq b\}}{\trlfplqbll{\Delta}{\multi{y}}{b}}
      {\trlfplqbll{B}{\sump{i}{y}}{b}}
    }
  }
  $$
  where $\theta$ and $\xi$ can be obtained from $\trlfplqbllpr{\rho}$ and $\trlfplqbllpr{\sigma}$,
  respectively, by Proposition~\ref{prop:strengthcon}.
\item
  If the last rule in $\pi$ is $E_\otimes$ and the immediate premises of $\pi$ are $\rho$ and $\sigma$, then
  $\trlfplqbllpr{\pi}$ is  
  {\footnotesize
  $$
  \infer[U]
  {\judgebll{\{\sump{i}{x}+\sump{i}{y}\leq b,1\leq b\}}{\trlfplqbll{\Gamma}{\multi{x}}{b},\trlfplqbll{\Delta}{\multi{y}}{b}}
        {\trlfplqbll{C}{\sump{i}{x}+\sump{i}{y}}{b}}}
  {
     \theta
     &
     \infer[L_{\exists x}]
     {
       \judgebll{\{\sump{i}{x}+\sump{i}{y}\leq b,1\leq b\}}{\trlfplqbll{\Gamma}{\multi{x}}{b},\trlfplqbll{A\otimes B}{\sump{i}{y}}{b}}
       {\trlfplqbll{C}{\sump{i}{x}+\sump{i}{y}}{b}}
     }
     {
       \infer[L_\otimes]
       {
         \judgebll{\{\sump{i}{x}+\sump{i}{y}\leq b,1\leq b,w+z\leq \sump{i}{y}\}}{\trlfplqbll{\Gamma}{\multi{x}}{b},\trlfplqbll{A}{w}{b}\otimes\trlfplqbll{B}{z}{b}}
         {\trlfplqbll{C}{\sump{i}{x}+\sump{i}{y}}{b}}
       }
       {
         \xi:\judgebll{\{\sump{i}{x}+\sump{i}{y}\leq b,1\leq b,w+z\leq \sump{i}{y}\}}{\trlfplqbll{\Gamma}{\multi{x}}{b},\trlfplqbll{A}{w}{b},\trlfplqbll{B}{z}{b}}
         {\trlfplqbll{C}{\sump{i}{x}+\sump{i}{y}}{b}}
       }
     }
  }
  $$}
  where $\xi$ can be obtained from $\trlfplqbllpr{\rho}$ by Proposition~\ref{prop:strengthcon} and Proposition~\ref{prop:monotonicity}
  and
  $$
  \theta:\judgebll{\{\sump{i}{x}+\sump{i}{y}\leq b,1\leq b\}}{\trlfplqbll{\Delta}{\multi{y}}{b}}{\trlfplqbll{A\otimes B}{\sump{i}{y}}{b}}
  $$
  can be obtained from $\trlfplqbllpr{\sigma}$ by Proposition~\ref{prop:strengthcon}.
\item 
  If the last rule in $\pi$ is $T$ and the immediate premises of $\pi$ are $\rho$ and $\sigma$, then
  $\trlfplqbllpr{\pi}$ is    
  $$
  \infer=[U]
  {
    \judgebll{\{x\leq b,1\leq b\}}{\trlfplqbll{\nat}{x}{b}}
    {\trlfplqbll{A}{x}{b}}
  }
  {
    \infer[P_!]
    {
      \judgebll{\{x\leq b,1\leq b\}}{}{!_{y<x}(\trlfplqbll{A}{y}{b}\linear \trlfplqbll{A}{y+1}{b})}
    }
    {
      \infer=[L_{\forall x},A,U]
      {
        \judgebll{\{y+1\leq b,1\leq b\}}{}{\trlfplqbll{A}{y}{b}\linear \trlfplqbll{A}{y+1}{b}}
      }
      {
        \infer[U]
        {
        \judgebll{\{y+1\leq b,1\leq b\}}{}{\trlfplqbll{A\linear A}{1}{b}}
        }
        {
          \sigma:\judgebll{\{y+1\leq b,1\leq b\}}{}{\trlfplqbll{\diamond}{1}{b}}
          &
          \theta:\judgebll{\{y+1\leq b,1\leq b\}}{\trlfplqbll{\diamond}{1}{b}}{\trlfplqbll{A\linear A}{1}{b}}
        }
      }
    }
    &
    \xi:\judgebll{\{x\leq b,1\leq b\}}{}{\trlfplqbll{A}{0}{b}}
    &
    \pi_x^A
  }
  $$
  where $\theta$ and $\xi$ can be obtained from $\trlfplqbllpr{\rho}$ and $\trlfplqbllpr{\sigma}$,
  respectively, by Proposition~\ref{prop:strengthcon} and $\sigma$ can be easily built.
\end{enumerate}
This concludes the proof.\qed

\begin{prop}\label{prop:wclfpl}
The correspondence $\trlfplqbllpr{\cdot}$ is weakly compositional. 
\end{prop}
\proof
A quick inspection on the proof of Theorem~\ref{thm:lfpl} shows that 
$\trlfplqbllpr{\pi}$ \emph{cannot} be obtained uniformly from 
$\trlfplqbllpr{\rho_1},\ldots,\trlfplqbllpr{\rho_1}$ (where $\rho_1,\ldots,
\rho_n$ are the immediate sub-proofs of $\pi$), because results like
Proposition~\ref{prop:strengthcon} or Proposition~\ref{prop:monotonicity} are often applied to
$\trlfplqbllpr{\rho_1},\ldots,\trlfplqbllpr{\rho_n}$ before they are
plugged together (in a uniform way) to obtain $\trlfplqbllpr{\pi}$.
All the results in Section~\ref{sect:properties}, however, transform
proofs to proofs preserving the underlying \Gtwoi\ proof. As a consequence
the embedding is weakly compositional.\qed

One may ask whether such an embedding might work for \BLL\ proper. We
believe this to be unlikely for several reasons. In particular, it 
seems that \BLL\ lacks a mechanism for turning the information
about the size of the manipulated objects from being global to being local. In \QBAL, this
r\^ole is played by first order quantifiers. 
As an example, consider the split function for lists of natural 
numbers that splits a list into two lists, one containing the 
even entries and one containing the odd entries. The type of that function in 
\LFPL\ is $\textbf{L}(\nat)\linear \textbf{L}(\nat)\otimes \textbf{L}(\nat)$ 
where $\textbf{L}(\cdot)$ denotes the type of lists that we have elided from our formal 
treatment for the sake of simplicity. In \QBAL\ this function gets the type 
$$
\textbf{L}_x(\nat_y)\linear \exists(u,v):
\{u+v\leq x\}.\textbf{L}_u(\nat_y)\otimes \textbf{L}_v(\nat_y).
$$
The only conceivable \BLL\ formula for this function is 
$\textbf{L}_x(\nat_y)\linear \textbf{L}_x(\nat_y)\otimes \textbf{L}_x(\nat_y)$. 
In \LFPL\ and in \QBAL\  we can compose the split function with
``append'' yielding a function of type $\textbf{L}_x(\nat_y)\linear \textbf{L}_x(\nat_y)$ that
can be iterated. 
In \BLL\ this composition receives the type $\textbf{L}_x(\nat_y)\linear
\textbf{L}_{2x}(\nat_y)$ which of course is not allowed in an iteration. 
But a hypothetical compositional embedding of \LFPL\ into \BLL\ would
have to be able to mimic this construction.
\section{Embedding \RRW}\label{sect:emrrw}
\noindent Ramified recurrence on words (\RRW) is a function algebra extensionally corresponding to polynomial time
functions introduced by Leivant in the early nineties~\cite{Leivant93}. Bellantoni and Cook's algebra \BC\ can
be easily embedded into \RRW.

Let $\wa$ be a word algebra, let $c^1_\wa,\ldots,c^w_\wa$ be the unary constructors of
$\wa$ and let $\zero_\wa$ be the only $0$-ary constructor of $\wa$.
$\id$ denotes the identity function on $\wa$. If $m$ is a natural
number and $1\leq i\leq m$, $\pi_i^m$ denotes the $i$-th projection on $m$ arguments in $\wa$. 
Given a $n$-ary function $g$ on $\wa$ and $n$ 
$m$-ary functions $f_1,\ldots,f_n$ on $\wa$, we can define
the $m$-ary \emph{composition} of $g$ and $f_1,\ldots,f_n$, denoted 
$\comp{g,f_1,\ldots,f_n}$, as follows:
$$
\comp{g,f_1,\ldots,f_n}(t_1,\ldots,t_m)=g(f_1(t_1,\ldots,t_m),\ldots,f_n(t_1,\ldots,t_m)).
$$
Given an $n$-ary function $f_{\zero_\wa}$ on $\wa$ and $n+2$-ary
functions $f_{c_\wa^1},\ldots,f_{c_\wa^w}$ on $\wa$, we can define
an $n+1$-ary function $g$, denoted $\rec{f_{c_\wa^1},\ldots,f_{c_\wa^w},f_{\zero_\wa}}$, 
by \emph{primitive recursion} as follows:
\begin{eqnarray*}
g(t_1,\ldots,t_n,\zero_\wa)&=&f_{\zero_\wa}(t_1,\ldots,t_n)\\
g(t_1,\ldots,t_n,c_\wa^i(t))&=&f_{c_\wa^i}(t_1,\ldots,t_n,g(t_1,\ldots,t_n,t),t)
\end{eqnarray*}
Given an $n$-ary function $f_{\zero_\wa}$ on $\wa$ and $n+1$-ary
functions $f_{c_\wa^1},\ldots,f_{c_\wa^w}$ on $\wa$, we can define
an $n+1$-ary function $g$, denoted $\cond{f_{c_\wa^1},\ldots,f_{c_\wa^w},f_{\zero_\wa}}$, 
as a \emph{conditional} as follows:
\begin{eqnarray*}
g(t_1,\ldots,t_n,\zero_\wa)&=&f_{\zero_\wa}(t_1,\ldots,t_n)\\
g(t_1,\ldots,t_n,c_\wa^i(t))&=&f_{c_\wa^i}(t_1,\ldots,t_n,t)
\end{eqnarray*}
We can generate functions starting from $\id$, $\pi_i^m$,
$\zero_\wa$ and $c_\wa$ by freely applying composition, primitive recursion and
conditional.

Not every function obtained this way is in \RRW: indeed, they correspond
to primitive recursive functions on $\wa$. In Figure~\ref{figure:rrw}, a formal
system for judgements in the form
$\vdash f:\pwa{i_1}\times\ldots\times\pwa{i_n}\rightarrow\pwa{i}$
(where $i_1,\ldots,i_n,i$ are natural numbers)
is defined. If such a judgement can be derived from the rules in Figure~\ref{figure:rrw},
then $f$ is said to be an \RRW\ function (the definition of \RRW\ given here is slightly
different but essentially equivalent to the original one~\cite{Leivant93}). Leivant~\cite{Leivant93} 
proved that \RRW\ functions are
exactly the polytime computable functions on $\wa$. But \RRW\ can be compositionally embedded into \QBAL, at
least in a weak sense.
\begin{figure*}[htbp]
\begin{center}
\fbox{
\begin{minipage}{.97\textwidth}
\centering\textbf{}
$$
\begin{array}{ccccccc}
\infer[]
{\vdash\id:\pwa{i}\rightarrow\pwa{i}}
{}
&\!\!&
\infer[]
{\vdash\zero_\wa:\pwa{i}}
{}
&\!\!&
\infer[]
{\vdash c_\wa:\pwa{i}\rightarrow\pwa{i}}
{}
&\!\!&
\infer[]
{\vdash\pi_i^m:\pwa{j_1}\times\ldots\times\pwa{j_m}\rightarrow\pwa{j}}
{j_i=j}
\end{array}
$$
$$
\begin{array}{c}
\infer[]
{\vdash\comp{g,f_1,\ldots,f_n}:\pwa{j_1}\times\ldots\pwa{j_m}\rightarrow\pwa{i}}
{
  \begin{array}{c}
    \vdash g:\pwa{i_1}\times\ldots\times\pwa{i_n}\rightarrow\pwa{i}\\
    \forall k\in\{1,\ldots,w\}.\vdash f_k:\pwa{j_1}\times\ldots\times\pwa{j_m}\rightarrow\pwa{i_k}
  \end{array}
}
\end{array}
$$
$$
\begin{array}{c}
\infer[]
{\vdash\rec{f_{c_\wa^1},\ldots,f_{c_\wa^w},f_{\zero_\wa}}:\pwa{i_1}\times\ldots\times\pwa{i_n}\times\pwa{j}\rightarrow\pwa{i}}
{
  \begin{array}{c}
    \forall k\in\{1,\ldots,w\}.\vdash f_{c_\wa^k}:\pwa{i_1}\times\ldots\times\pwa{i_n}\times\pwa{i}\times\pwa{j}\rightarrow\pwa{i}\\ 
    \vdash f_\zero:\pwa{i_1}\times\ldots\times\pwa{i_n}\rightarrow\pwa{i}
  \end{array}
  &
  i<j
}
\end{array}
$$
$$
\begin{array}{c}
\infer[]
{\vdash\cond{f_{c_\wa^1},\ldots,f_{c_\wa^w},f_{\zero_\wa}}:\pwa{i_1}\times\ldots\times\pwa{i_n}\times\pwa{j}\rightarrow\pwa{i}}
{
  \begin{array}{c}
    \forall k\in\{1,\ldots,w\}.\vdash f_{c_\wa^k}:\pwa{i_1}\times\ldots\times\pwa{i_n}\times\pwa{j}\rightarrow\pwa{i}\\
    \vdash f_\zero:\pwa{i_1}\times\ldots\times\pwa{i_n}\rightarrow\pwa{i}
  \end{array}
}
\end{array}
$$
\end{minipage}}
\caption{\RRW\ as a formal system.}
\label{figure:rrw}
\end{center}
\end{figure*}
Before embarking in the proof of that, however, we need a preliminary result.
\begin{lem}[Contraction Lemma]\label{lemma:contraction}
Every word algebra is duplicable, i.e., for every
word algebra $\wa$ there is a proof $\pi_\wa$ of
$\judgebll{}{\Wordalg_x}{\Wordalg_x\otimes\Wordalg_x}$ such
that $\sem{\pi_\wa}(x)=(x,x)$.
\end{lem}
\proof
For simplicity, consider the algebra $\N$ of natural numbers.
The proof we are looking for is the following:
{\scriptsize
$$
\infer=[U]
{\judgebll{}{\bN_x}{\bN_x\otimes\bN_x}}
{
  \infer[P_!]
  {\judgebll{}{}{!_{y<x}(\bN_y\otimes\bN_y\linear\bN_{y+1}\otimes\bN_{y+1})}}
  {
    \infer=[R_\otimes,L_\otimes,R_\linear]
    {
      \judgebll{}{}{\bN_y\otimes\bN_y\linear\bN_{y+1}\otimes\bN_{y+1}}
    }
    {
      \sigma:\judgebll{}{\bN_y}{\bN_{y+1}} & \sigma:\judgebll{}{\bN_y}{\bN_{y+1}}
    }
  }
  &
  \infer[R_\otimes]
  {\judgebll{}{}{\bN_{0}\otimes\bN_{0}}}
  {
    \rho:\judgebll{}{}{\bN_{0}} & \rho:\judgebll{}{}{\bN_{0}}
  }
  &
  \pi_x^{\bN_x\otimes\bN_x}
}
$$}
where $\sigma$ are $\rho$ are the proofs corresponding to $0$ and the successor
coming from \BLL. This concludes the proof.\qed

The following is the main result of this section:
\begin{thm}\label{theo:rrw}
\RRW\ can be embedded into \QBAL. Suppose, in other words,
that
$$
\pi:\vdash f:\pwa{i_1}\times\ldots\times\pwa{i_n}\rightarrow\pwa{i}
$$
and that $i< i_{j_1},\ldots,i_{j_m}$, while
$i=i_{k_1},\ldots,i_{k_h}$.
Then, there exist a \QBAL\ proof $\pi$ and a resource polynomial $q$ such that 
$$
\trrrwqbllpr{\pi}:
\judgebll{\{x_{k_1}\leq x,\ldots,x_{k_h}\leq x\}}
{\Wordalg_{x_1},\ldots,\Wordalg_{x_n}}
{\Wordalg_{q(x_{j_1},\ldots,x_{j_m})+x}}.
$$ 
where $(\prod_{i=1}^n\psi_\wa)\circ\Mean{\pi}\circ\varphi_\wa=f$. 
\end{thm}
\proof
By induction on the proof of
$$
\vdash f:\pwa{i_1}\times\ldots\times\pwa{i_n}\rightarrow\pwa{i}.
$$
Some interesting cases:
\begin{enumerate}[$\bullet$]
\item
  Consider the identity function $\id$. Clearly:
  $$
  \infer[]{\judgebll{\{x_1\leq x\}}{\Wordalg_{x_1}}{\Wordalg_{x}}}{}
  $$
\item
  Suppose $f=\comp{g,f_1,\ldots,f_n}$ and
  $$
  \infer[]
  {\vdash\comp{g,f_1,\ldots,f_n}:\pwa{j_1}\times\ldots\pwa{j_m}\rightarrow\pwa{i}}
  {\vdash g:\pwa{i_1}\times\ldots\times\pwa{i_n}\rightarrow\pwa{i} & \vdash f_k:\pwa{j_1}\times\ldots\times\pwa{j_m}\rightarrow\pwa{i_k}}
  $$
  We partition the sequence $i_1,\ldots,i_n$ into three sequences containing elements which
  are equal to $i$, strictly greater than $i$ and strictly smaller than $i$, respectively:
  $$
  \begin{array}{c}
    i_{k_1},\ldots,i_{k_h}\\
    i_{u_1},\ldots,i_{u_v}\\
    i_{a_1},\ldots,i_{a_b}\\
  \end{array}
  $$
  Clearly, $n=h+v+b$.
  Similarly for the sequence $j_1,\ldots,j_m$:
  $$
  \begin{array}{c}
    j_{t_1},\ldots,j_{t_e}\\
    j_{c_1},\ldots,j_{c_l}\\
    j_{z_1},\ldots,j_{z_d}\\
  \end{array}
  $$
  Again, $m=e+d+l$. 
  By induction hypothesis, there are proofs $\pi_g,\pi_{f_1},\ldots,\pi_{f_n}$ with the appropriate conclusions. 
  Now, consider the proofs $\pi_{f_{k_1}},\ldots,\pi_{f_{k_h}}$: they are the ones such that $i=i_{k_1},\ldots,i_{k_h}$. 
  By Proposition~\ref{prop:monotonicity}, we can
  assume that their conclusion is exactly the same, i.e., the polynomials $r_{k_1},\ldots,r_{k_h}$ in
  the rhs are indeed the same polynomial $r$. In other words, we have proofs
  \begin{eqnarray*}
    \rho_{f_{k_1}}&:&\judgebll{x_{t_1}\leq x,\ldots,x_{t_e}\leq x}{\Wordalg_{x_1},\ldots,\Wordalg_{x_m}}
    {\Wordalg_{q_{k_1}(x_1,\ldots,x_m,x)}}\\
    &\vdots& \\
    \rho_{f_{k_h}}&:&\judgebll{x_{t_1}\leq x,\ldots,x_{t_e}\leq x}{\Wordalg_{x_1},\ldots,\Wordalg_{x_m}}
    {\Wordalg_{q_{k_h}(x_1,\ldots,x_m,x)}}
  \end{eqnarray*}
  where $q_{k_0}(x_1,\ldots,x_m,x)=r(x_{z_1},\ldots,x_{z_d})+x$ for each $o$.
  The proofs $\pi_{f_{u_1}},\ldots,\pi_{f_{u_v}}$
  are the ones such that $i<i_{u_1},\ldots,i_{u_v}$. Observe
  that $x_{t_1},\ldots,x_{t_e}$ do not appear in the right hand side of their conclusions.
  By Proposition~\ref{prop:substpol} and Proposition~\ref{prop:strengthcon}, we can obtain proofs:
  \begin{eqnarray*}
    \rho_{f_{u_1}}&:&\judgebll{x_{t_1}\leq x,\ldots,x_{t_e}\leq x}{\Wordalg_{x_1},\ldots,\Wordalg_{x_m}}
    {\Wordalg_{q_{u_1}(x_1,\ldots,x_m,x)}}\\
    &\vdots& \\
    \rho_{f_{u_v}}&:&\judgebll{x_{t_1}\leq x,\ldots,x_{t_e}\leq x}{\Wordalg_{x_1},\ldots,\Wordalg_{x_m}}
    {\Wordalg_{q_{u_v}(x_1,\ldots,x_m,x)}}
  \end{eqnarray*}
  where $q_{u_1},\ldots,q_{u_v}$ only depend on $x_{z_1},\ldots,x_{z_d}$.
  Similarly, we can obtain proofs
  \begin{eqnarray*}
    \rho_{f_{a_1}}&:&\judgebll{x_{t_1}\leq x,\ldots,x_{t_e}\leq x}{\Wordalg_{x_1},\ldots,\Wordalg_{x_m}}
    {\Wordalg_{q_{a_1}(x_1,\ldots,x_m,x)}}\\
    &\vdots& \\
    \rho_{f_{a_b}}&:&\judgebll{x_{t_1}\leq x,\ldots,x_{t_e}\leq x}{\Wordalg_{x_1},\ldots,\Wordalg_{x_m}}
    {\Wordalg_{q_{a_b}(x_1,\ldots,x_m,x)}}    
  \end{eqnarray*}
  corresponding to $\pi_{f_{a_1}},\ldots,\pi_{f_{a_b}}$.
  Consider the proof $\pi_g$. By induction hypothesis it is a proof of
  $$
  \judgebll{y_{k_1}\leq y,\ldots,y_{k_h}\leq y}{\Wordalg_{y_1},\ldots,\Wordalg_{y_n}}{\Wordalg_{p(y_{u_1},\ldots,y_{u_v})+y}}
  $$
  Now, define a substitution mapping $y_i$ to $q_i(x_1,\ldots,x_m,x)$ and 
  $y$ to $r(x_{z_1},\ldots,x_{z_d})+x$.
  By Proposition~\ref{prop:substpol}, we can obtain a proof
  $$
  \judgebll{\consetone}{\Wordalg_{q_1(x_1,\ldots,x_m,x)},\ldots,\Wordalg_{q_n(x_1,\ldots,x_m,x)}}
  {\Wordalg_{s(y_{u_1},\ldots,y_{u_v})+y}}
  $$
  where
  $$
  \consetone=\{q_{k_1}(x_1,\ldots,x_m,x)\leq r(x_{z_1},\ldots,x_{z_d})+x,\ldots,q_{k_h}(x_1,\ldots,x_m,x)\leq r(x_{z_1},\ldots,x_{z_d})+x\}.
  $$
  But $\consetone$ is always true, and as a consequence
  $$
  \{x_{t_1}\leq x,\ldots,x_{t_e}\leq x\}\models\consetone.
  $$
  By Proposition~\ref{prop:strengthcon}, we can obtain a proof 
  $$
  \rho_g:\judgebll{x_{t_1}\leq x,\ldots,x_{t_e}\leq x}{\Wordalg_{q_1(x_1,\ldots,x_m,x)},\ldots,
    \Wordalg_{q_n(x_1,\ldots,x_m,x)}}{\Wordalg_{s(x_{z_1},\ldots,x_{z_d})+x}}
  $$
  Plugging the obtained proofs and remembering that base types are duplicable, we can construct a proof corresponding
  to $f=\comp{g,f_1,\ldots,f_n}$:
  $$
  \infer=
  {\judgebll{x_{t_1}\leq x,\ldots,x_{t_e}\leq x}{\Wordalg_{x_1},\ldots,\Wordalg_{x_m}}{\Wordalg_{p(x_{z_1},\ldots,x_{z_d})+x}}}
  {\rho_g & \rho_{f_1} & \ldots \rho_{f_n}}
  $$
\item
  Suppose $f=\rec{f_{c_\wa^1},\ldots,f_{c_\wa^w},f_{\zero_\wa}}$ and
  $$
  \infer[]
  {\vdash\rec{f_{c_\wa^1},\ldots,f_{c_\wa^w},f_{\zero_\wa}}:\pwa{i_1}\times\ldots\times\pwa{i_n}\times\pwa{j}\rightarrow\pwa{i}}
  {\vdash f_{c_\wa^k}:\pwa{i_1}\times\ldots\times\pwa{i_n}\times\pwa{i}\times\pwa{j}\rightarrow\pwa{i} 
    & \vdash f_\zero:\pwa{i_1}\times\ldots\times\pwa{i_n}\rightarrow\pwa{i} & i<j}
  $$
  By induction hypothesis, there are proofs $\pi_\zero$ and $\pi_{\wa}^1,\ldots,\pi_{\wa}^w$ with the appropriate conclusions.
  By Proposition~\ref{prop:monotonicity} we can assume that:
  \begin{eqnarray*}
    \pi_\zero&:&\judgebll{\consetone}{\Wordalg_{x_1},\ldots,\Wordalg_{x_n}}
      {\Wordalg_{q(x_{j_1},\ldots,x_{j_m},0)+x}}\\
    \pi_1&:&\judgebll{\consetone\cup\{x_{n+1}\leq x\}}{\Wordalg_{x_1},\ldots,\Wordalg_{x_{n+2}}}
      {\Wordalg_{q(x_{j_1},\ldots,x_{j_m},x_{n+2})+x}}\\
    &\vdots&\\
    \pi_w&:&\judgebll{\consetone\cup\{x_{n+1}\leq x\}}{\Wordalg_{x_1},\ldots,\Wordalg_{x_{n+2}}}
      {\Wordalg_{q(x_{j_1},\ldots,x_{j_m},x_{n+2})+x}}
  \end{eqnarray*} 
  where $\consetone=\{x_{k_1}\leq x,\ldots,x_{k_h}\leq x\}$ and $q$ is a fixed resource polynomial. Applying the
  substitution $\mu$ defined as 
  \begin{eqnarray*}
    x&\mapsto& (y+1)q(x_{j_1},\ldots,x_{j_m},y+1)+x\\
    x_{n+1}&\mapsto& (y+1)q(x_{j_1},\ldots,x_{j_m},y)+x\\
    x_{n+2}&\mapsto& y
  \end{eqnarray*}
  to $\pi_1,\ldots,\pi_w$ and by using Proposition~\ref{prop:monotonicity}, we obtain
  \begin{eqnarray*}
    \rho_1&:&\judgebll{\consettwo}{\Wordalg_{x_1},\ldots,\Wordalg_{x_n},\Wordalg_{(y+1)q(x_{j_1},\ldots,x_{j_m},y)+x},\Wordalg_{y}}
      {\Wordalg_{(y+2)q(x_{j_1},\ldots,x_{j_m},y+1)+x}}\\
    &\vdots&\\
    \rho_w&:&\judgebll{\consettwo}{\Wordalg_{x_1},\ldots,\Wordalg_{x_n},\Wordalg_{(y+1)q(x_{j_1},\ldots,x_{j_m},y)+x},\Wordalg_{y}}
      {\Wordalg_{(y+2)q(x_{j_1},\ldots,x_{j_m},y+1)+x}}
  \end{eqnarray*}
  where $\consettwo=\consetone[\sigma]\cup\{(y+1)q(x_{j_1},\ldots,x_{j_m},y)+x\leq (y+1)q(x_{j_1},\ldots,x_{j_m},y+1)+x\}$.
  Now, notice that 
  $$
  (y+1)q(x_{j_1},\ldots,x_{j_m},y)+x\leq (y+1)q(x_{j_1},\ldots,x_{j_m},y+1)+x
  $$
  always holds because $q$ is monotone. Moreover, $\consetone\models\consetone[\sigma]$.
  By Proposition~\ref{prop:strengthcon}, we can then replace $\consettwo$ with $\consetone$ in $\rho_1\ldots,\rho_w$.
  For simplicity, we use the following abbreviations:
  \begin{eqnarray*}
    A&=&\Wordalg_{x_1}\otimes\ldots\otimes\Wordalg_{x_n}\\
    B&=&\Wordalg_{x_1}\linear\ldots\linear\Wordalg_{x_n}\linear(A\otimes\Wordalg_{(y+1)q(x_{j_1},\ldots,x_{j_m},y)+x}\otimes\Wordalg_y)\\
    \Delta&=&\Wordalg_{x_1},\ldots,\Wordalg_{x_n}
  \end{eqnarray*}
  Now, from $\pi_\zero$ and from a proof corresponding to $\zero_\wa$ (with conclusion $\judgebll{\consetone}{}{\Wordalg_0}$), we construct
  $\sigma_\zero$ as follows:
  {\scriptsize
  $$
  \infer=[R_\linear]
  {\judgebll{\consetone}{}{B[y\leftarrow 0]}}
  {
    \infer=[\mbox{Lemma~\ref{lemma:contraction}}]
    {\judgebll{\consetone}{\Delta}{A\otimes\Wordalg_{q(x_{j_1},\ldots,x_{j_m},0)+x}\otimes\Wordalg_0}}
    {
      \infer=[R_\otimes]
      {\judgebll{\consetone}{\Delta,\Delta}
        {A\otimes\Wordalg_{q(x_{j_1},\ldots,x_{j_m},0)+x}\otimes\Wordalg_0}}
      {
        \infer[]
        {
          \judgebll{\consetone}{\Wordalg_{x_1}}{\Wordalg_{x_1}}
        }{}        
        &
        \cdots
        &
        \infer[]
        {
          \judgebll{\consetone}{\Wordalg_{x_n}}{\Wordalg_{x_n}}
        }{}        
        &
        \pi_\zero:\judgebll{\consetone}{\Delta}{\Wordalg_{q(x_{j_1},\ldots,x_{j_m},0)+x}}
        &
        \infer[]
        {
          \judgebll{\consetone}{}{\Wordalg_0}
        }{}
    }
  }}
  $$}\vspace{-6 pt}

  where we have used several times Lemma~\ref{lemma:contraction}.
  Similarly, from $\rho_i$ and from a proof corresponding to $c_\wa^i$ (with conclusion $\judgebll{\consetone}{\Wordalg_y}{\Wordalg_{y+1}}$),
  we construct $\sigma_i$ as follows:
  {\scriptsize
  $$
  \infer=[R_\linear]
  {\judgebll{\consetone}{}{B\linear B[y\leftarrow y+1]}}
    {\infer=[\mbox{Lemma~\ref{lemma:contraction}}]
      {\judgebll{\consetone}
        {B,\Delta}{A\otimes\Wordalg_{(y+2)q(x_{j_1},\ldots,x_{j_m},y+1)+x}\otimes\Wordalg_{y+1}}}
      {
        \infer=[L_\otimes,\mbox{Lemma~\ref{lemma:contraction}}]
        {\judgebll{\consetone}
          {B,\Delta,\Delta}{A\otimes\Wordalg_{(y+2)q(x_{j_1},\ldots,x_{j_m},y+1)+x}\otimes\Wordalg_{y+1}}}
        {
          \infer=[R_\otimes]
          {\judgebll{\consetone}
            {A,\Wordalg_{(y+1)q(x_{j_1},\ldots,x_{j_m},y)+x},\Wordalg_{y},\Wordalg_{y},\Delta}
            {A\otimes\Wordalg_{(y+2)q(x_{j_1},\ldots,x_{j_m},y+1)+x}\otimes\Wordalg_{y+1}}}
          {
            \infer[]        
            {
              \judgebll{\consetone}{A}{A}
            }{}        
            &
            \rho_i:\judgebll{\consetone}{\Delta,\Wordalg_{(y+1)q(x_{j_1},\ldots,x_{j_m},y)+x},\Wordalg_{y}}{\Wordalg_{(y+2)q(x_{j_1},\ldots,x_{j_m},y+1)+x}}
            &
            \infer[]
            {
              \judgebll{\consetone}{\Wordalg_{y}}{\Wordalg_{y+1}}
            }{}
          }
        }
      }
    }
$$}\vspace{-6 pt}

where, again, we have used several times
Lemma~\ref{lemma:contraction}.  And now we are ready to iterate over
the step functions: {\scriptsize
$$
\infer=[U]
{\judgebll{\consetone}{\Delta,\Wordalg_{x_{n+1}}}{\Wordalg_{(x_{n+1}+1)q(x_{j_1},\ldots,x_{j_m},x_{n+1})+x}}}
{
  \infer=[L_\linear,A,U]
  {\judgebll{\consetone}{\Wordalg_{x_{n+1}}}{B[y\leftarrow x_{n+1}]}}
  {
      \sigma_\zero
      &
      \sigma_1
      &
      \ldots
      &
      \sigma_w
  }
  &
  \infer=[A,R_\otimes,L_\linear]
  {\judgebll{\consetone}{\Delta,B[y\leftarrow x_{n+1}]}{\Wordalg_{(x_{n+1}+1)q(x_{j_1},\ldots,x_{j_m},x_{n+1})+x}}}
  {}
}
$$}
\item
  Suppose $f=\cond{f_{c_\wa^1},\ldots,f_{c_\wa^w},f_{\zero_\wa}}$ and
  $$
  \infer[]
  {\vdash\cond{f_{c_\wa^1},\ldots,f_{c_\wa^w},f_{\zero_\wa}}:\pwa{i_1}\times\ldots\times\pwa{i_n}\times\pwa{j}\rightarrow\pwa{i}}
  {\vdash f_{c_\wa}:\pwa{i_1}\times\ldots\times\pwa{i_n}\times\pwa{j}\rightarrow\pwa{i} 
    & \vdash f_\zero:\pwa{i_1}\times\ldots\times\pwa{i_n}\rightarrow\pwa{i}}
  $$
  We can distinguish three subcases:
  \begin{enumerate}[$\bullet$]
  \item
    If $i<j$, there are proofs $\pi_\zero$ and $\pi_{\wa}^1,\ldots,\pi_{\wa}^w$ with the appropriate conclusions.
    By Proposition~\ref{prop:monotonicity} we can assume that:
    \begin{eqnarray*}
      \pi_\zero&:&\judgebll{\consetone}{\Wordalg_{x_1},\ldots,\Wordalg_{x_n}}
      {\Wordalg_{q(x_{j_1},\ldots,x_{j_m},0)+x}}\\
      \pi_1&:&\judgebll{\consetone}{\Wordalg_{x_1},\ldots,\Wordalg_{x_{n+1}}}
      {\Wordalg_{q(x_{j_1},\ldots,x_{j_m},x_{n+1})+x}}\\
      &\vdots&\\
      \pi_w&:&\judgebll{\consetone}{\Wordalg_{x_1},\ldots,\Wordalg_{x_{n+1}}}
      {\Wordalg_{q(x_{j_1},\ldots,x_{j_m},x_{n+1})+x}}
    \end{eqnarray*} 
    where $\consetone=\{x_{k_1}\leq x,\ldots,x_{k_h}\leq x\}$. Applying the substitution
    $x_{n+1}\mapsto y$ to $\pi_1,\ldots,\pi_w$ and by using Proposition~\ref{prop:monotonicity} and Proposition~\ref{prop:strengthcon}, we obtain
    \begin{eqnarray*}
      \rho_1&:&\judgebll{\consetone}{\Wordalg_{x_1},\ldots,\Wordalg_{x_n},\Wordalg_{y}}
      {\Wordalg_{q(x_{j_1},\ldots,x_{j_m},y+1)+x}}\\
      &\vdots&\\
      \rho_w&:&\judgebll{\consetone}{\Wordalg_{x_1},\ldots,\Wordalg_{x_n},\Wordalg_{y}}
      {\Wordalg_{q(x_{j_1},\ldots,x_{j_m},y+1)+x}}
    \end{eqnarray*}
    For simplicity, wee use the following abbreviations:
    \begin{eqnarray*}
      A&=&\Wordalg_{x_1}\otimes\ldots\otimes\Wordalg_{x_n}\\
      B&=&\Wordalg_{x_1}\linear\ldots\linear\Wordalg_{x_n}\linear(A\otimes\Wordalg_{q(x_{j_1},\ldots,x_{j_m},y)+x}\otimes\Wordalg_y)\\
      \Delta&=&\Wordalg_{x_1},\ldots,\Wordalg_{x_n}
    \end{eqnarray*}
    Now, from $\pi_\zero$ and from a proof corresponding to $\zero_\wa$ (with conclusion $\judgebll{\consetone}{}{\Wordalg_0}$), we construct
    $\sigma_\zero$ as follows:
    {\scriptsize
      $$
      \infer=[R_\linear]
      {\judgebll{\consetone}{}{B[y\leftarrow 0]}}
      {
        \infer=[\mbox{Lemma~\ref{lemma:contraction}}]
        {\judgebll{\consetone}{\Delta}{A\otimes\Wordalg_{q(x_{j_1},\ldots,x_{j_m},1)+x}\otimes\Wordalg_0}}
        {
          \infer=[R_\otimes]
          {\judgebll{\consetone}{\Delta,\Delta}
            {A\otimes\Wordalg_{q(x_{j_1},\ldots,x_{j_m},1)+x}\otimes\Wordalg_0}}
          {
            \infer[]
            {
              \judgebll{\consetone}{\Wordalg_{x_1}}{\Wordalg_{x_1}}
            }{}        
            &
            \cdots
            &
            \infer[]
            {
              \judgebll{\consetone}{\Wordalg_{x_n}}{\Wordalg_{x_n}}
            }{}        
            &
            \pi_\zero:\judgebll{\consetone}{\Delta}{\Wordalg_{q(x_{j_1},\ldots,x_{j_m},1)+x}}
            &
            \infer[]
            {
              \judgebll{\consetone}{}{\Wordalg_0}
            }{}
          }
        }}
      $$}\vspace{-6 pt}

    Similarly, from $\rho_i$ and from a proof corresponding to
    $c_\wa^i$ (with conclusion
    $\judgebll{\consetone}{\Wordalg_y}{\Wordalg_{y+1}}$), we construct
    $\sigma_i$ as follows: {\scriptsize
      $$
      \infer=[R_\linear]
      {\judgebll{\consetone}{}{B\linear B[y\leftarrow y+1]}}
      {\infer=[\mbox{Lemma~\ref{lemma:contraction}}]
        {\judgebll{\consetone}
          {B,\Delta}{A\otimes\Wordalg_{q(x_{j_1},\ldots,x_{j_m},y+1)+x}\otimes\Wordalg_{y+1}}}
        {
          \infer=[L_\linear,A,U]
          {\judgebll{\consetone}
            {B,\Delta,\Delta}{A\otimes\Wordalg_{q(x_{j_1},\ldots,x_{j_m},y+1)+x}\otimes\Wordalg_{y+1}}}
          {
            \infer=[R_\otimes,W]
            {\judgebll{\consetone}
              {A,\Wordalg_{q(x_{j_1},\ldots,x_{j_m},y)+x},\Delta,\Wordalg_{y},\Wordalg_{y}}
              {A\otimes\Wordalg_{q(x_{j_1},\ldots,x_{j_m},y+1)+x}\otimes\Wordalg_{y+1}}}
            {
              \infer[]        
              {
                \judgebll{\consetone}{A}{A}
              }{}        
              &
              \rho_i:\judgebll{\consetone}{\Delta,\Wordalg_{y}}
              {\Wordalg_{q(x_{j_1},\ldots,x_{j_m},y+1)+x}}
              &
              \infer[]
              {
                \judgebll{\consetone}{\Wordalg_{y}}{\Wordalg_{y+1}}
              }{}
            }
          }
        }
      }
      $$}
    And now we are ready to iterate over the step functions:
    $$
    \infer=[U]
    {\judgebll{\consetone}{\Delta,\Wordalg_{x_{n+1}}}{\Wordalg_{q(x_{j_1},\ldots,x_{j_m},x_{n+1})+x}}}
    {
      \infer=[L_\linear,A,U]
      {\judgebll{\consetone}{\Wordalg_{x_{n+1}}}{B[y\leftarrow x_{n+1}]}}
      {
          \sigma_\zero
          &
          \sigma_1
          &
          \ldots
          &
          \sigma_w
      }
      &
      \infer=
      {\judgebll{\consetone}{\Delta,B[y\leftarrow x_{n+1}]}{\Wordalg_{q(x_{j_1},\ldots,x_{j_m},x_{n+1})+x}}}
      {}
    }
    $$
  \item
    If $i=j$, there are proofs $\pi_\zero$ and $\pi_{\wa}^1,\ldots,\pi_{\wa}^w$ with the appropriate conclusions.
    By Proposition~\ref{prop:monotonicity} we can assume that:
    \begin{eqnarray*}
      \pi_\zero&:&\judgebll{\consetone}{\Wordalg_{x_1},\ldots,\Wordalg_{x_n}}
      {\Wordalg_{q(x_{j_1},\ldots,x_{j_m},0)+x}}\\
      \pi_1&:&\judgebll{\consettwo}{\Wordalg_{x_1},\ldots,\Wordalg_{x_{n+1}}}
      {\Wordalg_{q(x_{j_1},\ldots,x_{j_m})+x}}\\
      &\vdots&\\
      \pi_w&:&\judgebll{\consettwo}{\Wordalg_{x_1},\ldots,\Wordalg_{x_{n+1}}}
      {\Wordalg_{q(x_{j_1},\ldots,x_{j_m})+x}}
    \end{eqnarray*} 
    where $\consetone=\{x_{k_1}\leq x,\ldots,x_{k_h}\leq x\}$ and
    $\consettwo=\consetone\cup\{x_{n+1}\leq x\}$. 
    Applying the substitution
    $x_{n+1}\mapsto y$ to $\pi_1,\ldots,\pi_w$ and by using Proposition~\ref{prop:substpol}, we obtain
    \begin{eqnarray*}
      \rho_1&:&\judgebll{\consetthree}{\Wordalg_{x_1},\ldots,\Wordalg_{x_n},\Wordalg_{y}}
      {\Wordalg_{q(x_{j_1},\ldots,x_{j_m})+x}}\\
      &\vdots&\\
      \rho_w&:&\judgebll{\consetthree}{\Wordalg_{x_1},\ldots,\Wordalg_{x_n},\Wordalg_{y}}
      {\Wordalg_{q(x_{j_1},\ldots,x_{j_m})+x}}
    \end{eqnarray*}
    where $\consetthree=\consetone\cup\{y\leq x\}$
    For simplicity, we use the following abbreviations:
    \begin{eqnarray*}
      A&=&\Wordalg_{x_1}\otimes\ldots\otimes\Wordalg_{x_n}\\
      B&=&\Wordalg_{x_1}\linear\ldots\linear\Wordalg_{x_n}\linear(A\otimes\Wordalg_{q(x_{j_1},\ldots,x_{j_m})+x}\otimes\Wordalg_y)\\
      \Delta&=&\Wordalg_{x_1},\ldots,\Wordalg_{x_n}
    \end{eqnarray*}
    Now, from $\pi_\zero$ and from a proof corresponding to $\zero_\wa$ (with conclusion $\judgebll{\consetthree}{}{\Wordalg_0}$), we construct
    $\sigma_\zero$ as follows:
    {\scriptsize
      $$
      \infer=[R_\linear]
      {\judgebll{\consetthree}{}{B[y\leftarrow 0]}}
      {
        \infer=[\mbox{Lemma~\ref{lemma:contraction}}]
        {\judgebll{\consetthree}{\Delta}{A\otimes\Wordalg_{q(x_{j_1},\ldots,x_{j_m})+x}\otimes\Wordalg_0}}
        {
          \infer=[R_\otimes]
          {\judgebll{\consetthree}{\Delta,\Delta}
            {A\otimes\Wordalg_{q(x_{j_1},\ldots,x_{j_m})+x}\otimes\Wordalg_0}}
          {
            \infer[]
            {
              \judgebll{\consetthree}{\Wordalg_{x_1}}{\Wordalg_{x_1}}
            }{}        
            &
            \cdots
            &
            \infer[]
            {
              \judgebll{\consetthree}{\Wordalg_{x_n}}{\Wordalg_{x_n}}
            }{}        
            &
            \pi_\zero:\judgebll{\consetthree}{\Delta}{\Wordalg_{q(x_{j_1},\ldots,x_{j_m})+x}}
            &
            \infer[]
            {
              \judgebll{\consetthree}{}{\Wordalg_0}
            }{}
          }
        }}
      $$}\vspace{-6 pt}

    Similarly, from $\rho_i$ and from a proof corresponding to $c_\wa^i$ 
    (with conclusion $\judgebll{\consetthree}{\Wordalg_y}{\Wordalg_{y+1}}$),
    we construct $\sigma_i$ as follows:
    {\scriptsize
      $$
      \infer=[R_\linear]
      {\judgebll{\consetthree}{}{B\linear B[y\leftarrow y+1]}}
      {\infer=[\mbox{Lemma~\ref{lemma:contraction}}]
        {\judgebll{\consetthree}
          {B,\Delta}{A\otimes\Wordalg_{q(x_{j_1},\ldots,x_{j_m})+x}\otimes\Wordalg_{y+1}}}
        {
          \infer=[\mbox{Lemma~\ref{lemma:contraction}}]
          {\judgebll{\consetthree}
            {B,\Delta,\Delta}{A\otimes\Wordalg_{q(x_{j_1},\ldots,x_{j_m})+x}\otimes\Wordalg_{y+1}}}
          {
            \infer=[R_\otimes]
            {\judgebll{\consetthree}
              {A,\Wordalg_{q(x_{j_1},\ldots,x_{j_m})+x},\Wordalg_{y},\Wordalg_{y},\Delta}
              {A\otimes\Wordalg_{q(x_{j_1},\ldots,x_{j_m})+x}\otimes\Wordalg_{y+1}}}
            {
              \infer[]        
              {
                \judgebll{\consetthree}{A}{A}
              }{}        
              &
              \rho_i:\judgebll{\consetthree}{\Delta,\Wordalg_{y}}
              {\Wordalg_{q(x_{j_1},\ldots,x_{j_m})+x}}
              &
              \infer[]
              {
                \judgebll{\consetthree}{\Wordalg_{y}}{\Wordalg_{y+1}}
              }{}
            }
          }
        }
      }
      $$}\vspace{-6 pt}

    And now we are ready to iterate over the step functions:
    {\scriptsize
     $$
     \infer[U]
     {\judgebll{\consettwo}{\Delta,\Wordalg_{x_{n+1}}}{\Wordalg_{q(x_{j_1},\ldots,x_{j_m})+x}}}
     {
       \infer=
       {\judgebll{\consettwo}{\Wordalg_{x_{n+1}}}{B[y\leftarrow x_{n+1}]}}
       {
         \sigma_\zero:\judgebll{\consettwo}{}{B[y\leftarrow 0]}
         &
           \infer[]
           {\judgebll{\consettwo}{}{!_{y<x_{n+1}}(B\linear B[y\leftarrow y+1])}}
           {\sigma_i:\judgebll{\consetthree}{}{B\linear B[y\leftarrow y+1]}}
         }
       &
       \infer=
       {\judgebll{\consettwo}{\Delta,B[y\leftarrow x_{n+1}]}{\Wordalg_{q(x_{j_1},\ldots,x_{j_m})+x}}}
       {}
     }
     $$}
  \item
    If $i>j$, the proof is similar to the case $i<j$.
  \end{enumerate}
  Observe how, in all the three cases, the proof corresponding to $f$ is structurally the same.
\end{enumerate}
This concludes the proof.\qed

Quite interestingly, the proof of Theorem~\ref{theo:rrw} is very similar in structure to
the proof of polynomial time soundness for \BC\ given in~\cite{Bellantoni92}, which is based on the following observation:
the size of the output of a \BC\ function is bounded by a polynomial on the sizes of normal arguments plus the \emph{maximum}
of sizes of safe arguments. This cannot be formalized in \BLL, because the resource polynomials do not include any function
computing the maximum of its arguments. On the other hand, this can be captured in \QBAL\ by way of constraints.
\begin{prop}\label{prop:wcrrw}
The correspondence $\trrrwqbllpr{\cdot}$ is weakly compositional. 
\end{prop}
\proof The proof is essentially identical to the one of
Proposition~\ref{prop:wclfpl}.\qed

\section{Conclusions}
\noindent We presented \QBAL, a new ICC system embedding two distinct and unrelated systems for
impredicative recursion in the sense of \cite{hofmann:sigact}, namely ramified recurrence
and non-size increasing computation.
\QBAL\ allows to overcome the main weakness of \BLL, namely that
all resource variables are global. In the authors' view, this
constitutes the first step towards unifying ICC systems into
a single framework. The next step consists in defining an embedding
of light linear logic into \QBAL\ and the authors are currently
investigating on that.
\bibliographystyle{plain}
{\small \bibliography{bllr}}
\end{document}